**Ratnadip Adhikari**

**A Treatise on Stability of Autonomous and Non-autonomous Systems**

Ratnadip Adhikari

# A Treatise on Stability of Autonomous and Non-autonomous Systems

Theory and Illustrative Practical Applications

# ACKNOWLEDGEMENT

It was a pleasant and remarkable experience to write the present book. I am personally benefited a lot through this work from the viewpoint of both mathematical knowledge as well as writing skills. During the course of this work, I had learnt so many interesting and valuable concepts which have a lot of theoretical as well as practical significance.

The successful completion of the book could hardly be possible without the sincere assistance of a number of individuals. I will take this opportunity to express my cordial gratitude to all those people who helped me either directly or indirectly during this important work.

First of all I wish to express my sincere appreciation and due respect to **Dr. N. Sukavanam**, Associate Professor Department of Mathematics, IIT Roorkee for his continuous and able guidance, constant encouragements and positive motivations which have crucial significance in organizing the ideas and concepts of this book. He had always shown keen interests in my queries regarding this topic and gladly devoted his precious time whenever necessary.

I am also grateful to some of my friends and research scholars of Department of Mathematics, IIT Roorkee for their important suggestions and valuable advices which helped me a lot while preparing this book.

I owe a lot to my mother for her constant love and support. She always encouraged me to have positive and independent thinking which really matter in my life. I would like to thank her very much and share this moment of happiness with her.

Last but not the least, I am also thankful to many faculty members of the Department of Mathematics, IIT Roorkee for their unselfish help which I got whenever needed during the course of my work.

<div style="text-align: right;">**RATNADIP ADHIKARI**</div>

# ABSTRACT

Stability is a very important property of any physical system. By a stable system, we broadly mean that small disturbances either in the system inputs or in the initial conditions do not lead to large changes in the overall behavior of the system. To be of practical use, a system must have to be stable. The theory of stability is a vast, rapidly growing subject with prolific and innovative contributions from numerous researchers. As such, an introductory book that covers the basic concepts and minute details about this theory is essential. The primary aim of this book is to make the readers familiar with the various terminologies and methods related to the stability analysis of time-invariant (*autonomous*) and time-varying (*non-autonomous*) systems. A special treatment is given to the celebrated Liapunov's direct method which is so far the most widely used and perhaps the best method for determining the stability nature of both autonomous as well as non-autonomous systems. After discussing autonomous systems to a considerable extent, the book concentrates on the non-autonomous systems. From stability point of view, these systems are often quite difficult to manage. Also, unlike their autonomous counterparts, the non-autonomous systems often behave in peculiar manners which can make the analysts arrive at misleading conclusions. Due to these issues, this book attempts to present a careful and systematic study about the stability properties of non-autonomous systems.

The book in total consists of eight chapters which are organized as follows. The first chapter introduces the various terminologies and definitions, associated with stability theory. Stability of linear time-invariant systems is discussed in Chapter 2. Chapter 3 deals with the concept of *linearization* which is often used to analyze stability of some types of non-linear autonomous systems. The fourth chapter describes Liapunov's approach of stability analysis for autonomous systems. The fifth chapter introduces the notion of stability of time-invariant systems and demonstrates various associated pitfalls and peculiarities through numerous illustrative examples. Chapter 6 deals with the topic of exponential stability of non-autonomous systems which has significant practical importance. Liapunov's direct method for stability of time-invariant system is studied in Chapter 7. Finally, Chapter 8 provides an overview of stability of periodic and discrete time systems which are often encountered in real-world. A summarization of the topics covered in this book is provided as the epilogue.

It should be noted that although the book provides an intensive study of stability theory but this is in no way exhaustive. The theory of stability is a quite vast area and it is rationally impossible to cover all relevant topics in an introductory book like this one. However, my efforts will be rewarded and I shall feel very much satisfied if the book meets the expectations of the readers and they benefit through reading it.

# CONTENTS

# LIST OF FIGURES

*Chapter-1*

# INTRODUCTION

## 1.1 Basic Concepts of systems and Control

In most branches of Applied Mathematics, the aim is to analyze a given situation under certain predefined *simplifying assumptions*. For example if a mass is suspended from a fixed point by a string then the basic assumptions might be that the air resistance, mass of the string and dimensions of the body can all be neglected and also that the gravitational attraction is constant. Under such assumptions, a mathematical problem then will be to determine the nature of small motions of the mass about the equilibrium position. The word *system* is generally used to mean a collection of objects which are related by mutual interactions within a set of given or chosen entities. The interactions are among the entities inside the system those either influence or get influenced by the entities outside the system. The mathematical model which we develop for a system is based on the assumption that corresponding to each input to the given system there are in general a number of possible outputs.

In our everyday life we encounter with various systems, known as the *physical* or *dynamical* systems. In dealing with the physical systems, we mainly focus upon those which behave in some desired fashion, i.e. whose behaviors we wish to be under control. These are known as the *control systems*. Generally a set of differential or difference equations are used to formulate the mathematical model of a dynamical system. Yet another useful form of mathematical description of a dynamical system is the *state variable* approach. The state of a system is defined as the smallest set of variables which must have to be known at a given instant in order to be able to calculate the future response of the system to any specified input. The mathematical variables used to specify a certain state of a system are known as the state variables. The space containing all the state variables is called the *state space* of the system.

In general, it is possible to change the outputs of a control system in any prescribed fashion (which are at least within the reasonable limits) by means of intelligent manipulation of its inputs. Of course, the more realistic the model, the more difficult is to solve the resulting mathematical equations. Thus, the real control systems which are often quite complicated must have to be presented in a simplified form so that they can be mathematically dealt with ease.

Now-a-days the theory of control systems is applied to many areas which include the use of robots in automated manufacturing, automatic control of large-scale power systems, numerical control of machine tools, autopilots for aircrafts etc. In the following sections, we discuss about the common mathematical representation of a dynamical control system and define some associated concepts related to the state of a system.

## 1.2 Mathematical Formulation of A Dynamical System

The differential equation of a dynamical control system is given by:

$$\dot{\mathbf{X}}(t) = f(\mathbf{X}, t) \tag{1.1}$$

where,

$$\mathbf{X}(t) = \begin{bmatrix} x_1(t) \\ x_2(t) \\ \ldots \\ \ldots \\ \ldots \\ x_n(t) \end{bmatrix} \text{ and } f = \begin{bmatrix} f_1 \\ f_2 \\ \ldots \\ \ldots \\ \ldots \\ f_n \end{bmatrix} \text{ are } n \times 1 \text{ column vectors and each } f_i \text{ is a function of}$$

the *n* state variables $x_i$ (*i*=1, 2, 3,..., *n*) and time *t*. Also, $\dot{\mathbf{X}} = \dfrac{d\mathbf{X}}{dt}$.

We assume that the functions $f_i$ are continuous and satisfy the standard conditions so that the solution of the Eq. (1.1) exists and is unique for the given initial conditions. We can also see that it is not necessary to write the above equations separately; rather, they can be represented in a combined form as one or several equations of higher order.

## 1.3 Different Types of Dynamical Systems

There are different types of dynamical systems based on the forms of the Eq. (1.1). If in the Eq. (1.1), $f_i$ are linear functions then the system is called *linear* otherwise *nonlinear*. A linear control system can be always expressed in the form:

$$\dot{\mathbf{X}}(t) = \mathbf{A}(t)\mathbf{X}(t) + \mathbf{B}(t)\mathbf{U}(t) \tag{1.2}$$

where

$$\mathbf{X}(t) = \begin{bmatrix} x_1(t) \\ x_2(t) \\ \ldots \\ \ldots \\ \ldots \\ x_n(t) \end{bmatrix}, \mathbf{U}(t) = \begin{bmatrix} u_1(t) \\ u_2(t) \\ \ldots \\ \ldots \\ \ldots \\ u_p(t) \end{bmatrix} \text{ are respectively the } n \times 1 \text{ state vector with the } n$$

state variables $x_i$ (*i*=1, 2, 3,..., *n*) and $p \times 1$ input vector with the *p* input (control) variables $u_j$ (*j*=1, 2, 3,..., *p*). $\mathbf{A}(t)$ and $\mathbf{B}(t)$ are respectively the $n \times n$ and $n \times p$ real matrices. Since, the state vector $\mathbf{X} \in R^n$ and the input vector $\mathbf{U} \in R^p$, so $R^n$ is the *n*-dimensional state space, whereas $R^p$ is the *p*-dimensional input space. For specific values of the state vector and the input vector, the system represented by Eq. (1.2) will result in definite outputs.

Eq. (1.2) is also sometimes known as the *state equation*.

If the Eq. (1.1) depends explicitly on time *t*, it is known as **Non-autonomous** or **Time-variant**, otherwise **Autonomous** or **Time-invariant**. In an autonomous system, none of the functions $f_i$ explicitly depends on time *t*. Thus, in this case the matrices **A**(*t*) and **B**(*t*) are constant matrices, independent of time *t*.

## 1.4 Concept of Stability of A Dynamical System

Problems of stability appeared for the first time in Mechanics during the investigation of an equilibrium state of a system. The criterion for stability of rigid bodies in equilibrium under gravitational forces was first formulated by E. Torricelli in 1644. Later in 1788, G. Lagrange proved a sufficient condition for stability of equilibrium of any conservative system. The basic idea of stability analysis in Applied Mathematics is to deal with the various aspects of a given system and its responses to small deviations under certain pre-defined simplifying assumptions. Roughly speaking, by stability of a dynamical system we mean that:

> *The small changes in system inputs or initial conditions do not result in large changes in the overall system behavior.*

To be of practical use, any control system which we encounter with must have to be stable. It is relatively easy to study the stability of linear autonomous systems. There are several powerful tests, such as the **Routh-Hurwitz criterion** which yields necessary and sufficient conditions for stability of such systems. Also in a stable linear time-invariant system with given bounded inputs, the output is always bounded. But, there is no such general method to study the stability of a nonlinear system. Moreover, in case of a nonlinear system, with bounded inputs, the resulting output may not be bounded. But almost all the common dynamical systems have certain unavoidable nonlinearities associated with them. So, it is necessary to make an in depth study about the stability criteria of nonlinear systems.

Next, when we turn to non-autonomous systems, the situation becomes much more complicated. There are no robust mathematical techniques to determine the nature of stability for such a system. We shall see that how the things get fundamentally changed from the stability analysis perspective if the system changes with time. After going into further details, we shall find that non-autonomous systems are much complex and different from autonomous systems in various ways. Thus, unlike autonomous case it is not so easy to study the time varying systems. But, these systems of course have great practical significance.

In this book, we shall at first develop the stability theories for linear and nonlinear autonomous systems. Our particular focus will be on the celebrated Liapunov's stability theory. Then, we shall study a number of well known methods to determine the stability properties of time varying systems. Some important types of stability such as exponential stability, uniform stability, etc. will be considered. After a reasonable extent of study of continuous systems, we shall provide an overview of the discrete time-varying systems. The important case of periodic systems will also be considered. All concepts of stability analysis in this book will be illustrated through various important practical examples.

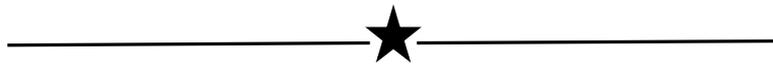

*Chapter-2*

# STABILITY OF LINEAR AUTONOMOUS SYSTEMS

## 2.1 Equilibrium Point of A Dynamical System

Let us consider the system represented by the differential equation:

$$\dot{\mathbf{X}}(t) = f(\mathbf{X}, t) \tag{2.1}$$

where, $\mathbf{X}(t) = \begin{bmatrix} x_1(t) \\ x_2(t) \\ \vdots \\ \vdots \\ \vdots \\ x_n(t) \end{bmatrix}$ is the $n \times 1$ state vector and $f = \begin{bmatrix} f_1(t) \\ f_2(t) \\ \vdots \\ \vdots \\ f_n(t) \end{bmatrix}$ is the $n \times 1$ column vector with components $f_i(x_1, x_2, ..., x_n, t)$ ($i=1,2,3,...,n$).

If $f(\mathbf{C}, t) = 0 \; \forall \; t$ where $\mathbf{C}$ is some constant vector, then it follows from Eq. (2.1) that if $\mathbf{X}(t_0) = \mathbf{C}$, then $\mathbf{X}(t) = \mathbf{C} \; \forall \; t \geq t_o$ so that all the solutions starting at $\mathbf{C}$ will remain there. Here the vector $\mathbf{C} \in R^n$ is said to be an **equilibrium point** or a **critical point** of the system, represented by Eq. (2.1).

Now, we consider the linear autonomous system with the state equation:

$$\dot{\mathbf{X}}(t) = \mathbf{A}\mathbf{X}(t) + \mathbf{B}\mathbf{U}(t) \tag{2.2}$$

where, $\mathbf{X}$ is the $n \times 1$ state vector, $\mathbf{U}$ is the $p \times 1$ input vector and $\mathbf{A}$, $\mathbf{B}$ are respectively the $n \times n$ and $n \times p$ real constant matrices.

If, for any constant input vector $\mathbf{U}(t) = \mathbf{U}^e$, $\exists$ a point $\mathbf{X}(t) = \mathbf{X}^e$ such that $\dot{\mathbf{X}}(t) = 0 \; \forall \; t$, then the point $\mathbf{X}^e$ is called an equilibrium point of the system represented by Eq. (2.2). Now, from Eq. (2.2), we have: $\mathbf{A}\mathbf{X}^e + \mathbf{B}\mathbf{U}^e = 0$ which gives

$$\mathbf{X}^e = -\mathbf{A}^{-1}\mathbf{B}\mathbf{U}^e \tag{2.3}$$

If, the matrix $\mathbf{A}$ is singular then from above it is clear that $\mathbf{X}^e$ is not a *discrete point* but a *continuum of points*.

By a proper translation in the form of $\mathbf{X} = \mathbf{X}^e + \mathbf{C}$, the equilibrium point $\mathbf{X}^e$ can always be transferred to the origin $\mathbf{X} = 0$, i.e. in this case we have from Eq. (2.1), $f(0, t) = 0 \; \forall \; t$. Thus, we can always take the origin as the equilibrium point of a system. We shall also assume that the origin is an *isolated equilibrium point*, i.e. there is no other constant solution of the system in the neighborhood (nbd.) of the origin.

The solution **X**(*t*) of the Eq. (2.1), regarded as a function of *t* in the *n*-dimensional state space is called a *trajectory*. The equilibrium point **X**=0 (the $n\times 1$ zero vector) is a common point of all trajectories of the system.

## 2.2 Concept of Stability in the Sense of Liapunov

There is no universal single concept of stability and it is defined in many different ways. But, we shall consider the following fundamental definition of stability due to a Russian mathematician Liapunov:

An equilibrium state **X**=0 is said to be:

- *Stable*, if for any $\varepsilon>0$, $\exists$ a real number $\delta(\varepsilon)>0$, such that $\|\mathbf{X}(t_0)\|\leq\delta \Rightarrow \|\mathbf{X}(t)\|\leq\varepsilon$, $\forall t\geq t_0$.
- *Asymptotically stable*, if it is stable and in addition to that $\mathbf{X}(t)\to 0$ as the time $t\to\infty$.
- *Monotonically stable*, if it is asymptotically stable and the distance of the state from the origin decreases monotonically with time.
- *Globally asymptotically stable*, if it is stable in the Liapunov sense and $\mathbf{X}(t)\to 0$ as $t\to\infty$, corresponding to every initial state $\mathbf{X}(t_0)$.
- *Unstable*, if it is not stable, i.e., if there is some $\varepsilon>0$, so that $\forall \delta>0$, $\|\mathbf{X}(t_0)\|<\delta \Rightarrow \|\mathbf{X}(t_1)\|>\varepsilon$ for some $t_1>t_0$. If this holds for every $t_0$ satisfying $\|\mathbf{X}(t_0)\|<\delta$, then the equilibrium is said to be *completely unstable*.

Here, $\|.\|$ means a vector norm. Generally we use the Euclidean norm, which is given by:

$$\|\mathbf{X}\|=(x_1^2+x_2^2+\ldots+x_n^2)^{1/2}.$$

The intuitive idea of stability of a dynamical system is that for small deviations from the equilibrium state, the subsequent motions of the trajectories are not too large. To understand the concept of stability let us visualize it in a two dimensional system as depicted by the figures:

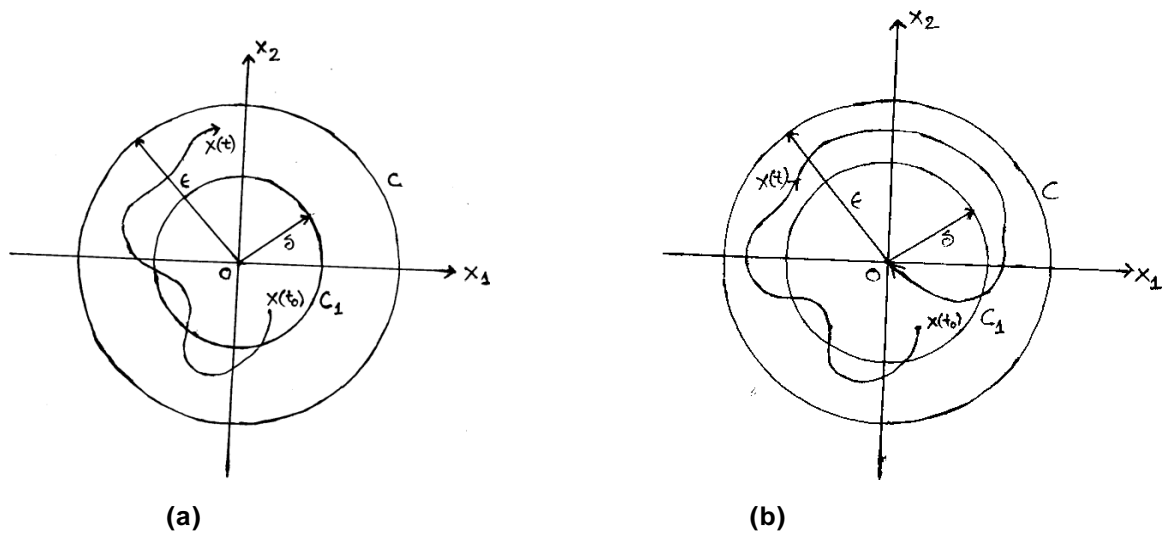

**(a)**          **(b)**

**Fig 2.1: (a) stable equilibrium, (b) asymptotic stable equilibrium**

Here, if the origin O is a stable equilibrium point, then for given any outer circle C of radius $\varepsilon$, there must exist an inner circle $C_1$ of radius $\delta$, such that the trajectories starting at a point $\mathbf{X}(t_0)$ inside the circle $C_1$ never leaves the circle C. Also if the origin O is asymptotically stable, then any trajectory of the system starting inside the circle $C_1$, tends towards the origin O, as $t \to \infty$.

### 2.2.1 A Physical Example of Stability

Let us consider a ball resting in equilibrium on a sheet of metal, bent into various shapes with cross-section as shown below.

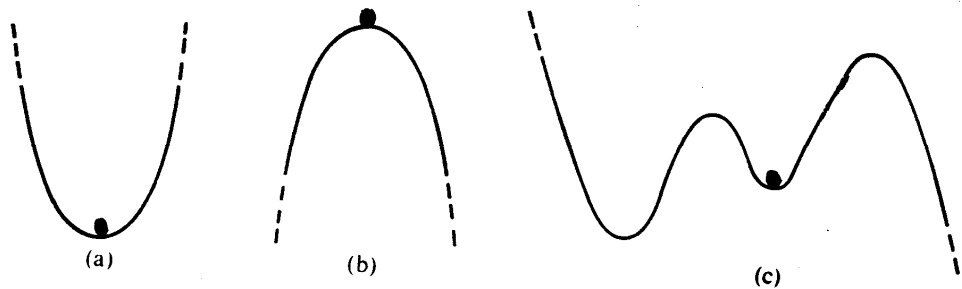

**Fig 2.2: Equilibrium of a ball resting on a sheet of metal**

If frictional forces are neglected, then small disturbances lead to:

(a) Oscillatory motion about the equilibrium point.
(b) The ball moving away without returning to equilibrium.
(c) Oscillatory motion about equilibrium, unless the initial disturbance is too large that the ball is forced to oscillate about the new equilibrium position on the left or to fall off at the right.

Here, case (a) represents stability, case (b) represents asymptotic stability and case (c) represents stability which is not globally asymptotical.

**NOTE: Some important remarks**

(a) Different equilibrium points have different stability properties.
(b) Only very small disturbances are allowable in stability analysis.
(c) Asymptotic stability is more desirable for practical purposes.

### 2.2.2 Classification of Critical Points

There are five types of critical points which are described below.

- **Improper node:** An *improper node* is a critical point P at which all the trajectories except for two of them have the same limiting direction of the tangent.
- **Proper node:** A *proper node* is a critical point P at which every trajectory has a definite limiting direction and for any direction at P, there is a trajectory having that direction as its limiting direction.

- **Saddle point:** A *saddle point* is a critical point P at which there are two incoming trajectories, two outgoing trajectories and all other trajectories in a nbd. of P bypass P.
- **Centre:** A *centre* is a critical point which is enclosed by infinitely many closed trajectories.
- **Spiral point:** A *spiral point* is a critical point P about which the trajectories spiral approaching P as time $t\to\infty$.

**Theorem 2.1:** Consider the general linear autonomous system given by $\dot{\mathbf{X}} = \mathbf{AX}$. Then the system is asymptotically stable at the origin if and only if **A** is a stability matrix, i.e. all the eigen values of the coefficient matrix **A** have negative real parts.

**Proof:** The autonomous linear system is given by

$$\dot{\mathbf{X}} = \mathbf{AX} \tag{2.4}$$

Here **X** is an $n \times 1$ column vector and **A** is the $n \times n$ coefficient matrix.

We assume that the matrix **A** has the distinct eigen values $\lambda_i$ ($i=1,2,3,...,n$) and $\mathbf{X}_i$ be the eigen vector of **A** corresponding to the eigen value $\lambda_i$, i.e. $\mathbf{AX}_i = \lambda_i \mathbf{X}_i$. Then, the solution of the system (2.3) is given by:

$$\mathbf{X}(t) = c_1 e^{\lambda_1 t} \mathbf{X}_1 + c_2 e^{\lambda_2 t} \mathbf{X}_2 + --------- c_n e^{\lambda_n t} \mathbf{X}_n \tag{2.5}$$

Where, $c_i$ ($i=1,2,3,...,n$) are constants. Thus, we have

$$\|\mathbf{X}(t)\| = \|c_1 e^{\lambda_1 t} \mathbf{X}_1 + c_2 e^{\lambda_2 t} \mathbf{X}_2 + --------- c_n e^{\lambda_n t} \mathbf{X}_n\|$$

$$\leq \sum_{i=1}^{n} |c_i e^{\lambda_i t}| \|\mathbf{X}_i\|$$

$$\leq \sum_{i=1}^{n} |c_i| |e^{\text{Re}(\lambda_i t)}| \|\mathbf{X}_i\| \tag{2.6}$$

Now if $\text{Re}(\lambda_i) < 0, \forall\ i$, then it is clear from (2.6) that $\|\mathbf{X}(t)\| \to 0$, as $t \to \infty$, i.e. $\mathbf{X}(t) \to 0$ as $t \to \infty$. Thus, in this case the system is *asymptotically stable* at the origin. Again if $\text{Re}(\lambda_k) > 0$, for any k, then from (2.5), we see that $\|\mathbf{X}(t)\| \to \infty$ as $t \to \infty$. So, in this case the system is *unstable* at the origin. Similar results also hold well when the eigen values of **A** are not distinct, i.e. some of the eigen values of the **A** are repeated.

*Hence, the theorem is proved*.

### NOTE: Important consequences of the theorem
- The system (2.4) is stable if $\text{Re}(\lambda_i) < 0$ for all *i*, unstable if any $\text{Re}(\lambda_i) > 0$ and completely unstable if all $\text{Re}(\lambda_i) > 0$.

- ➢ By this theorem, we have seen that the stability of a linear autonomous system can be directly determined by finding the eigen values of the coefficient matrix **A**.
- ➢ If the real part of some eigen value of **A** is zero, i.e. if Re($\lambda_i$)=0 for some *i*, then this theorem cannot be applied and in this case it is not easy to decide the stability of the system.

### 2.2.3 Bounded Input Bounded Output Stability

A control system is said to be *Bounded Input Bounded Output (BIBO)* stable if for any given bounded input the output of the system is also bounded. If the system outputs as well as all the state variables are bounded for all initial conditions for any bounded input, then the system is known as *totally stable*.

**Theorem 2.2:** Consider the state equation given by $\dot{\mathbf{X}}(t) = \mathbf{A}(t)\mathbf{X}(t) + \mathbf{B}(t)\mathbf{U}(t)$. Then, if the system is asymptotically stable, then it is also BIBO stable.

### NOTE: Important consequences of the theorem

- ➢ According to the Theorem 2.2, if the system, represented by the state equation $\dot{\mathbf{X}}(t) = \mathbf{A}(t)\mathbf{X}(t) + \mathbf{B}(t)\mathbf{U}(t)$ is asymptotically stable, then with bounded input, output is always bounded.
- ➢ From definition of total stability and the two theorems stated, it is clear that the input-output (I/O) system is totally stable, iff all the eigen values of the coefficient matrix **A** have negative real parts.

## 2.3 Some Illustrative Mathematical Examples

In this section, we present a number of mathematical examples to enhance the understanding of different types of stability, discussed so far.

**Example 2.1:** Let us consider the linear autonomous system $\dot{\mathbf{X}} = \mathbf{A}\mathbf{X}$, where the state vector $\mathbf{X} = \begin{bmatrix} x_1 \\ x_2 \end{bmatrix}$ and the coefficient matrix $\mathbf{A} = \begin{bmatrix} -3 & 1 \\ 1 & -3 \end{bmatrix}$. Now, the eigen values of **A** are $\lambda_1 = -2$ and $\lambda_2 = -4$. The corresponding eigen vectors are given by $\mathbf{A}_1 = \begin{bmatrix} 1 \\ 1 \end{bmatrix}$ and $\mathbf{A}_2 = \begin{bmatrix} 1 \\ -1 \end{bmatrix}$. Then the solution of the system is:

$$\mathbf{X}(t) = \begin{bmatrix} x_1 \\ x_2 \end{bmatrix} = c_1 \mathbf{A}_1 e^{\lambda_1 t} + c_2 \mathbf{A}_2 e^{\lambda_2 t}$$

$$= c_1 \begin{bmatrix} 1 \\ 1 \end{bmatrix} e^{-2t} + c_2 \begin{bmatrix} 1 \\ -1 \end{bmatrix} e^{-4t}.$$

Thus, the solution can be written as:
$x_1(t) = c_1 e^{-2t} + c_2 e^{-4t}$.
$x_2(t) = c_1 e^{-2t} - c_2 e^{-4t}$.

For equilibrium point, we put $\dot{X}(t) = 0$ which gives $X = 0$, i.e. the origin (as the coefficient matrix **A** is non-singular).

Thus we see that $X(t) = \begin{bmatrix} x_1 \\ x_2 \end{bmatrix} \to 0$, as $t \to \infty$. Hence, the system is **asymptotically stable** at the origin, which is also an **improper node** for the given system.

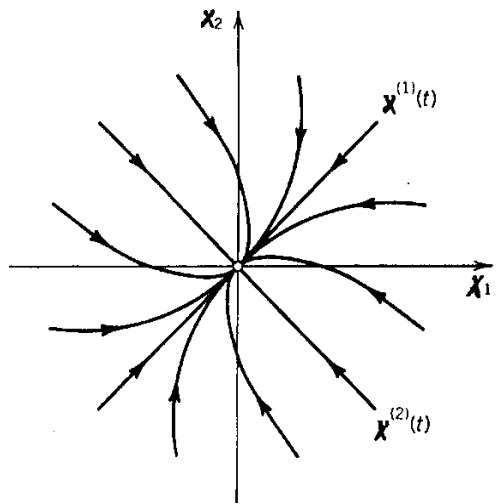

**Fig 2.3: Asymptotical stability at the origin (improper node)**

Here we see that the real parts of all the eigen values of **A** are negative. Thus, by applying the Theorem 2.1, we can directly see that the given system is asymptotically stable at the origin.

**Example 2.2:** Let us consider the system $\dot{X} = AX$, where $A = \begin{bmatrix} 1 & 0 \\ 0 & 1 \end{bmatrix}$. Here, the eigen values of **A** are 1, 1. Now, the solution of the system is given by, $x_1(t) = c_1 e^t$, $x_2(t) = c_2 e^t$ or $x_1 = kx_2$, $k$ being some constant. Taking $\dot{X}(t) = 0$, we see that the origin $X = 0$ is the only critical point of the system. Now from the solution of the system, we see that as $t \to \infty$, $x_1 \to \pm\infty$, $x_2 \to \pm\infty$, i.e. $X(t) \to \infty$. Thus the system is **unstable** at the origin, which is also a **proper node** in this case.

Here we see that all the eigen values of **A** are positive. Thus by Theorem 2.1, we can directly verify that the given system is unstable (actually completely unstable) at the origin. The trajectories of the system are depicted in Fig. 2.4.

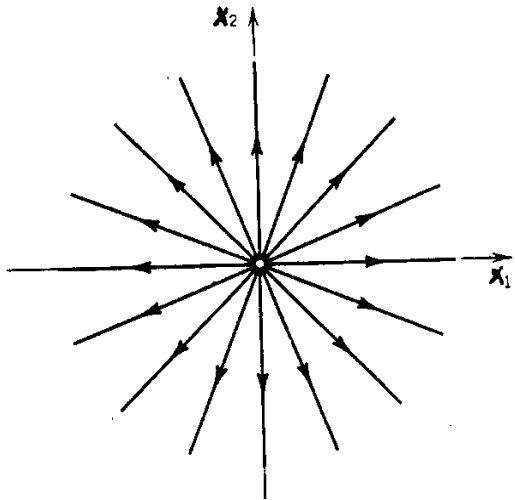

**Fig 2.4: Completely unstable critical point (proper node)**

**Example 2.3:** Let us consider the system $\dot{\mathbf{X}} = \mathbf{A}\mathbf{X} = \begin{bmatrix} 1 & 0 \\ 0 & -1 \end{bmatrix} \mathbf{X}$. Here the eigen values of **A** are 1,-1. Since the eigen value -1<0, so the system is *unstable* at the origin, which is the only critical point of the system. Here we can see that the solution of the system can be written as $x_1 x_2$=constant which represents a family of hyperbolas, as can be seen from the Fig. 2.5. From the figure, it is also clear that the origin is a saddle point for the given system.

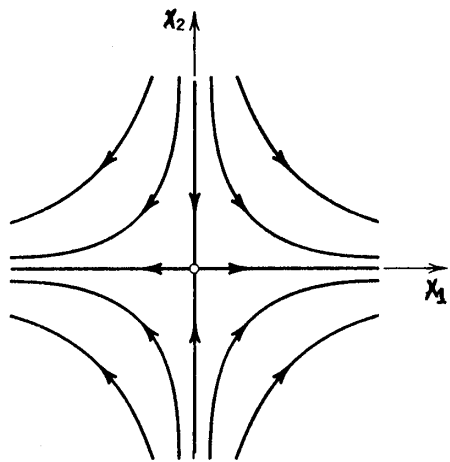

**Fig 2.5: Unstable equilibrium point (saddle point)**

**Example 2.4:** Let us consider the system $\dot{\mathbf{X}} = \mathbf{A}\mathbf{X} = \begin{bmatrix} 0 & 1 \\ -4 & 0 \end{bmatrix} \mathbf{X}$. Here the eigen values of **A** are $\pm 2i$. So the solution of the system is given by

$$x_1(t) = c_1 e^{2it} + c_2 e^{-2it}.$$
$$x_2(t) = 2i(c_1 e^{2it} - c_2 e^{-2it}).$$

Putting $\dot{\mathbf{X}} = 0$, we see that the only critical point is the origin (0,0). As both the eigen values of **A** are zero, so we cannot apply the Theorem 2.1 here to examine the stability. Now the solution of the system can be written as $x_1^2 + x_2^2/4 = $ constant which represents a family of concentric ellipses. Thus from definition, the system is **stable** at the origin, which is also a **centre** of the system in this case. The trajectories of the system are depicted in Fig. 2.6.

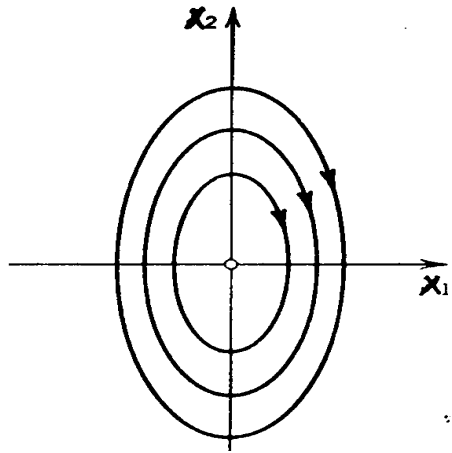

**Fig 2.6: Stability at the origin (centre)**

**Example 2.5:** Let us consider the system, $\dot{\mathbf{X}} = \mathbf{A}\mathbf{X} = \begin{bmatrix} -1 & 1 \\ -1 & -1 \end{bmatrix} \mathbf{X}$. Here the eigen values of **A** are $-1 \pm i$ and the only critical point is the origin. So we can easily verify that the system is **asymptotically stable** at the origin which is also a **spiral point** of the system.

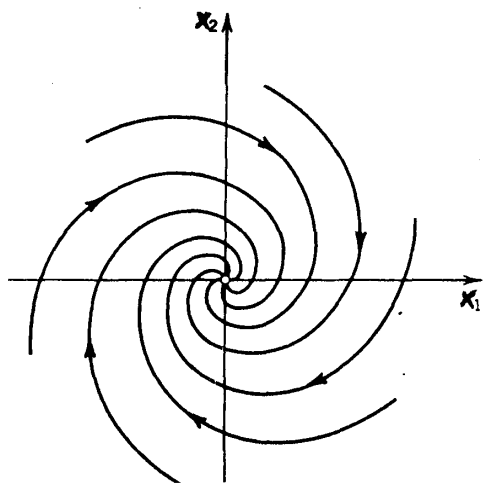

**Fig 2.7: Asymptotical stability at the origin (spiral point)**

**NOTE:** The above mathematical examples illustrate that for a linear autonomous system if we know the coefficient matrix A, then we can easily determine the nature of stability of the system by finding the eigen values of A.

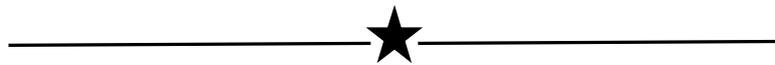

*Chapter-3*

# STUDY OF NONLINEAR AUTONOMOUS SYSTEMS

### 3.1 Nonlinear Dynamical System

We have so far discussed the stability properties of linear systems. However, most of the practical systems are nonlinear in nature which cannot be studied easily like linear systems. A nonlinear system differs from the usual linear system in a number of ways, like:

- *Principle of superposition* is not applicable to nonlinear systems.
- Altering the size of the input does not change the shape of the response of a linear system, but this is not in the case of a nonlinear system.
- In a nonlinear system, stability is a characteristic of the system, independent of the magnitude of the system input or the initial conditions. But, in a linear system stability may depend on the magnitude of the input as well as the initial conditions.

General study of the stability properties of a nonlinear system is a quite difficult job. Also unlike a linear system, the analytical solution of a nonlinear system is rarely possible.

### 3.2 Concept of Linearization

One of the well-known methods for studying the *local stability* (i.e. stability in the small) of an isolated equilibrium point for a nonlinear autonomous system is by *linearization* of the model, provided it is possible. We can linearize a nonlinear model only by restricting the system variables to sufficiently small deviations about an equilibrium point. The method of linearization of a nonlinear autonomous system is described below.

Linearization is based on the Taylor Series expansion of a nonlinear function about a critical point. The Taylor series expansion of a nonlinear function $f(x)$ is given by

$$f(x) = f(x_0) + \left(\frac{df}{dx}\right)_{x_0} (x - x_0) + \left(\frac{d^2 f}{dx^2}\right)_{x_0} \frac{(x - x_0)^2}{2!} + \cdots \qquad (3.1)$$

Now we can get a linear approximation of *f(x)* if we ignore all the terms on the Right Hand Side (RHS) except the first, provided that these are very small. We generalize this idea to the case of a nonlinear autonomous system, whose state equation is given by

$$\dot{\mathbf{X}} = f(\mathbf{X}, \mathbf{U}) \qquad (3.2)$$

where, **X** and *f* are both *n*×1 matrices.

Now, ignoring the higher order terms in the Taylor series expansion of *f*(**X**) and assuming that the equilibrium point is the origin, we get the following linearized form of the model (3.2).

$$\dot{\mathbf{X}} = \mathbf{A}\mathbf{X} + \mathbf{B}\mathbf{U} \tag{3.3}$$

where, $\mathbf{A} = \begin{bmatrix} \frac{\partial f_1}{\partial x_1} & \frac{\partial f_1}{\partial x_2} & - & - & \frac{\partial f_1}{\partial x_n} \\ \frac{\partial f_2}{\partial x_1} & \frac{\partial f_2}{\partial x_2} & - & - & \frac{\partial f_2}{\partial x_n} \\ - & - & - & - & - \\ - & - & - & - & - \\ \frac{\partial f_n}{\partial x_1} & \frac{\partial f_n}{\partial x_2} & - & - & \frac{\partial f_n}{\partial x_n} \end{bmatrix}$ and $\mathbf{B} = \begin{bmatrix} \frac{\partial f_1}{\partial u_1} & \frac{\partial f_1}{\partial u_2} & - & - & \frac{\partial f_1}{\partial u_n} \\ \frac{\partial f_2}{\partial u_1} & \frac{\partial f_2}{\partial u_2} & - & - & \frac{\partial f_2}{\partial u_n} \\ - & - & - & - & - \\ - & - & - & - & - \\ \frac{\partial f_n}{\partial u_1} & \frac{\partial f_n}{\partial u_2} & - & - & \frac{\partial f_n}{\partial u_n} \end{bmatrix}$ are called the *Jacobian*

*matrices*.

With the formulation described above, we can now investigate the local stability of the linearized model (3.3) by studying the matrix **A**. Also after applying the linearization to the system (3.2), we can treat the new system (3.3) as a linear system.

### 3.3 Limit Cycles in A Nonlinear System

The definition of stability in the sense of Liapunov includes the possibility that the state of a deviated system may follow a closed trajectory within the tolerance limit specified by the circle with radius $\varepsilon > 0$. This behavior of a nonlinear system is called a *limit cycle* and corresponds to an oscillation of fixed amplitude and period.

As an example, let us consider the behavior of an electronic oscillator which is described by the Van der Pol's differential equation

$$\frac{d^2 x}{dt^2} - \mu(1-x^2)\frac{dx}{dt} + x = 0 \tag{3.4}$$

This differential equation describes the electrical circuit as shown in the figure below:

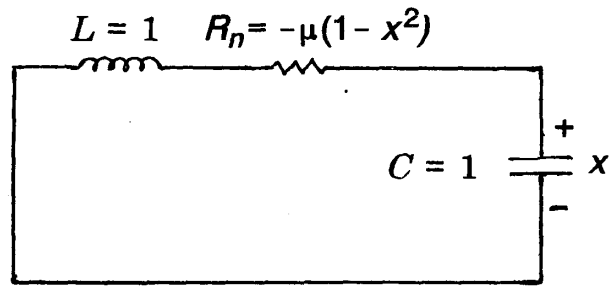

**Fig 3.1: Electrical circuit diagram of the Van der Pol's equation**

Here $R_n$ represents a nonlinear resistor and $x$ is the voltage across a capacitor. For small $x$, the resistance is negative. As $x$ increases, the resistance increases and is positive for $x>1$. For an intermediate value of $x$, i.e. for $0<x<1$, the oscillations are stable.

An important aim in the study of nonlinear systems is the determination of the existence and location of the limit cycles. In general, a limit cycle is undesirable in a control system, but in some cases it can be tolerated if the amplitude of oscillations is small.

### 3.4 Some Illustrative Mathematical examples

Here, we provide some practical examples to illustrate the concept of linearization.

**Example 3.1:** Consider the second order nonlinear differential equation

$$\frac{d^2x}{dt^2} + (\frac{dx}{dt})^2 + \frac{dx}{dt} + x(x-2) = 0 \tag{3.5}$$

To determine the equilibrium points and to examine the stability of the system at these points, first we take $x_1 = x$ and $x_2 = \dot{x}_1$. Then, the state equations for the system (3.5) are obtained as

$$\left. \begin{array}{l} f_1 = \dot{x}_1 = x_2 \\ f_2 = \dot{x}_2 = -\dot{x}_1{}^2 - x_2{}^2 + 2x_1 - x_2 \end{array} \right\} \tag{3.6}$$

Putting $\dot{\mathbf{X}} = 0$ in (3.6), the critical points are found as (0,0) and (2,0). Now applying the concept of linearization, we get a linear model for the given system as

$$\dot{\mathbf{X}} = \mathbf{A}\mathbf{X} \tag{3.7}$$

where, the Jacobian matrix **A** is given by

$$\mathbf{A} = \begin{bmatrix} \frac{\partial f_1}{\partial x_1} & \frac{\partial f_1}{\partial x_2} \\ \frac{\partial f_2}{\partial x_1} & \frac{\partial f_2}{\partial x_2} \end{bmatrix} = \begin{bmatrix} 0 & 1 \\ 2(1-x_1) & -(1+2x_2) \end{bmatrix}.$$

Now at (0,0) we have $\mathbf{A} = \begin{bmatrix} 0 & 1 \\ 2 & -1 \end{bmatrix}$. Eigen values of **A** are 1,-2. As, 1>0, so the given system is unstable in the nbd. of the critical point (0,0). In this case we say that the system is *locally unstable* at (0,0).

At the point (2,0) $\mathbf{A} = \begin{bmatrix} 0 & 1 \\ -2 & -1 \end{bmatrix}$. The two eigen values of **A** are ½(-1±i√7) both of which have negative real parts and so the system is *locally asymptotically stable* at (2,0).

**Example 3.2:** Let us consider the second-order nonlinear differential equation

$$\frac{d^2x}{dt^2} + x^2(\frac{dx}{dt} - 1) + x = 0 \tag{3.8}$$

Taking, $x_1 = x$ and $x_2 = \dot{x}_1$, the state equations are obtained as

$$\dot{\mathbf{X}} = \begin{bmatrix} f_1 \\ f_2 \end{bmatrix} = \begin{bmatrix} x_2 \\ x_1^2(1-x_2) - x_1 \end{bmatrix} \quad (3.9)$$

Putting $\dot{\mathbf{X}} = 0$ in (3.9), the critical points of the system are found as (0,0) and (1,0). Now applying the concept of linearization, we get the following linear model for the given system:

$$\dot{\mathbf{X}} = \mathbf{AX} = \begin{bmatrix} 0 & 1 \\ 2x_1(1-x_2) & -x_1^2 \end{bmatrix} \mathbf{X}. \quad (3.10)$$

At the critical point (0,0), we have $\mathbf{A} = \begin{bmatrix} 0 & 1 \\ -1 & 0 \end{bmatrix}$, which has the eigen values $\pm i$. Now the solution of the system (3.10) represents a family of concentric circles in the phase plane $x_1$-$x_2$. Thus, from definition, the system is *locally stable* at the origin, which is also a *centre* in this case. Also, the characteristic polynomial of A is $\lambda^2+1$, which indicates an oscillatory system and so we can expect a *limit cycle* in the nbd. of (0,0).

At the critical point (1,0), we have $\mathbf{A} = \begin{bmatrix} 0 & 1 \\ 1 & -1 \end{bmatrix}$. The two eigen values of **A** are ½(-1±√5), both of which are negative. Thus, the system is *locally asymptotically* stable at (1,0).

**Example 3.3:** Let us consider the system, which is represented by the equations

$$\left. \begin{array}{l} \dot{x}_1 = x_1 - 2x_1 x_2 \\ \dot{x}_2 = -2x_2 + x_1 x_2 \end{array} \right\} \quad (3.11)$$

This is an example of the **Lotka–Volterra Prey-predator model**, commonly used in biology where, $x_1(t)$ and $x_2(t)$ respectively represents the population of the prey and predator species at time *t*. Here, our linear model would be

$$\dot{\mathbf{X}} = \mathbf{AX} = \begin{bmatrix} 1-2x_2 & -2x_1 \\ x_2 & -2+x_1 \end{bmatrix} \mathbf{X}. \quad (3.12)$$

where, **A** is the Jacobian matrix.

Putting $\dot{\mathbf{X}} = 0$ in (3.12), the critical points are obtained as (0,0) and (2,½).

At the critical point (0,0), $\mathbf{A} = \begin{bmatrix} 1 & 0 \\ 0 & -2 \end{bmatrix}$ which has the eigen values 1,-2. Hence the system is *locally unstable* at (0,0).

At (2,½), $\mathbf{A} = \begin{bmatrix} 0 & -4 \\ \frac{1}{2} & 0 \end{bmatrix}$, whose eigen values are $\pm i\sqrt{2}$. The solution of the system can be obtained as $\frac{x_1^2}{2} + 4x_2^2 = $ constant, which represents a family of concentric ellipses. Thus, from definition, the given system is locally stable at the critical point (2,½).

**Example 3.4:** Let us consider a simple pendulum, consisting of a mass *m* and a rod of length *L*, suspended from a fixed point as shown below:

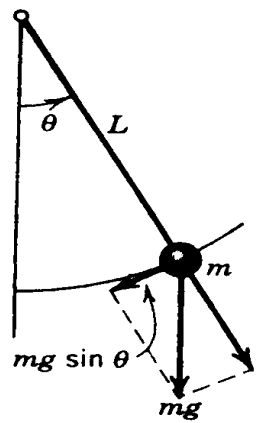

**Fig 3.2: Motion of a simple pendulum**

Let $\theta$ be the angular displacement and let the mass of the rod be negligible. Now the equation of motion of the simple pendulum is given by

$$mL\ddot{\theta} + mg \sin\theta = 0$$
$$\Rightarrow \ddot{\theta} + k \sin\theta = 0$$

where, $k = g/L$.

When $\theta$ is small, then $\sin\theta \to \theta$. Thus, the equation of motion becomes $\ddot{\theta} + k\theta = 0$. This has the solution

$$\theta = A\cos\sqrt{k}t + B\sin\sqrt{k}t \qquad (3.13)$$

where, *A*, *B* are arbitrary constants.

Now, we put $\theta = x_1$ and $\dot{\theta} = x_2$. Then from the equation of the simple pendulum, we get

$$\left.\begin{aligned} \dot{x}_1 &= x_2 \\ \dot{x}_2 &= -k \sin x_1 \end{aligned}\right\} \qquad (3.14)$$

Here, we have infinitely many critical points of the form ($n\pi$,0), where $n \in Z$.

For the critical point (0,0), we have $\sin\theta = \sin(x_1+0) \to x_1$, by linearization. Thus, our linear model becomes

$$\dot{\mathbf{X}} = \mathbf{AX} = \begin{bmatrix} 0 & 1 \\ -k & 0 \end{bmatrix} \mathbf{X}. \tag{3.15}$$

We can see that (0,0) is a *centre* for the system (3.15) and so is a point of *stable equilibrium*. As $\sin\theta$ is periodic of period $2\pi$, so all the critical points of the form ($n\pi$,0), ($n=0,\pm2, \pm4,...$) are centers and the system is *locally asymptotically stable* at these points.

Again for the critical point ($\pi$,0) we take $\sin\theta = \sin(x_1+\pi) \to -\sin x_1 \to -x_1$. Then our linear model becomes

$$\dot{\mathbf{X}} = \mathbf{AX} = \begin{bmatrix} 0 & 1 \\ k & 0 \end{bmatrix} \mathbf{X}. \tag{3.16}$$

Here, we can see that the system is *locally unstable* at the critical point ($\pi$,0) which is also a *saddle point* for the system. Thus, as before the system is *locally unstable* at all the critical points ($n\pi$,0), ($n=\pm1,\pm3,...$) which are also the *saddle points* of the system. Now we have from the state equations (3.14)

$$\frac{dx_2}{dx_1} = \frac{-k \sin x_1}{x_2}$$

which gives on integration, $x_2^2 = 2kx_1 + c$, where $c$ is an arbitrary constant to be determined from the given initial conditions.

The trajectories of the simple pendulum in the phase plane are as depicted below:

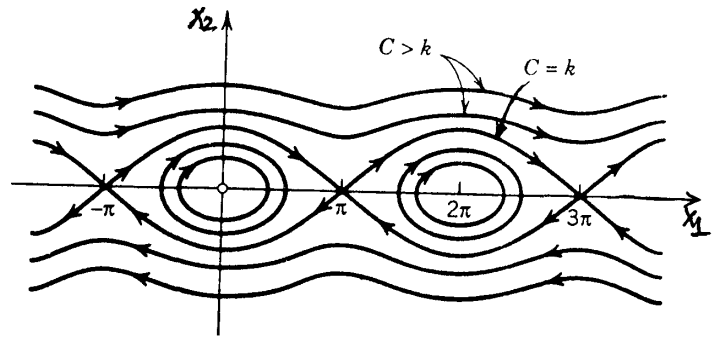

**Fig 3.3: Trajectories of a simple pendulum**

**Example 3.5:** Let us apply a *dumping force* $c\dot{\theta}$ to the equation of the simple pendulum. Then, the new equation of motion becomes

$$\ddot{\theta} + c\dot{\theta} + k\sin\theta = 0 \tag{3.17}$$

As usual, by applying linearization, we get the linear model as

$$\dot{\mathbf{X}} = \mathbf{AX} = \begin{bmatrix} 0 & 1 \\ -k & -c \end{bmatrix} \mathbf{X}. \tag{3.18}$$

Also the critical points of the system are $(n\pi, 0)$ where $n \in Z$.

Here we have two cases:
- If $c=0$, there is no damping and this case is same as the previous case of simple undamped pendulum.
- If $c>0$, there is damping. As before, here also, since $\sin\theta$ is a periodic function of period $2\pi$, so the stability natures of all the critical points of the form $(n\pi,0)$, $(n=0,\pm2,\pm4,...)$ are same as that of the *centre* $(0,0)$. Similarly, the stability natures of all the critical points of the form $(n\pi,0)$, $(n=\pm1,\pm3, ...)$ are same as that of the *saddle point* $(\pi,0)$.

The trajectories of the damped pendulum are shown below:

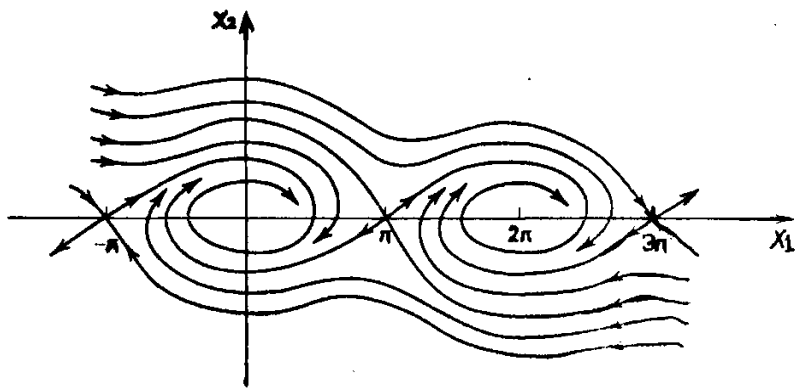

**Fig 3.4: Trajectories of the damped pendulum**

### NOTE: Concluding remarks

Linearization method is applied only for sufficiently small deviations from the equilibrium points, i.e. for the study of local stability of some autonomous nonlinear systems. However, in many cases the region of validity of the local stability of system is not known and so linearization becomes useless. In such cases, we use Liapunov's direct method of stability which is described in the next chapter.

*Chapter-4*

# LIAPUNOV'S STABILITY THEORY FOR AUTONOMOUS SYSTEMS

## 4.1 A Brief Introduction to Liapunov's Theory

This theory was introduced and further developed by the celebrated Russian mathematician A.M. Liapunov. Liapunov published his Ph.D. thesis entitled: ***The General Problem of Motion Stability*** in 1892. This monograph includes so many fruitful ideas and results of primary significance that the whole theory of stability can be divided into two periods, viz. *Pre-Liapunov* period and *Post-Liapunov* periods.

First of all Liapunov provided a rigorous definition of motion stability. The absence of such a definition had often caused misunderstandings since otherwise a motion which is stable in one sense can be unstable in another. Then he suggested two main methods for analyzing the stability problems of motions. Of these, the second method, also called *The Direct Liapunov Method* is widely popular due to its simplicity and efficiency. The formulae and results given by Liapunov are for stability of both autonomous (also called *steady state motions*) and non-autonomous systems. These results are effectively used to solve various problems of stability occurring in physical situations. In this chapter, we shall study the main methods and results as developed by Liapunov to determine the nature of stability of autonomous systems. Liapunov's theory for the stability properties of nonlinear autonomous systems mainly concerns systems of the form

$$\dot{\mathbf{X}} = f(\mathbf{X}) \tag{4.1}$$

where, $f(0) = 0$, $\mathbf{X}(t_0) = \mathbf{X}_0$. As usual, $\mathbf{X}$, $f$ are both $n \times 1$ column vectors and also $f(\mathbf{X})$ is a continuous function.

The simple stability criteria developed for linear systems are not applicable to the nonlinear systems. Also in many cases the information obtained from linearization of a nonlinear system becomes inconclusive. In such cases we use the Liapunov's direct method of stability which is based on the well-known result of Mechanics that states that:

> *In a conservative system, an equilibrium point is stable if the energy of the system is a minimum.*

The main aim of the method is to determine the stability nature of the equilibrium point at the origin of the system (4.1) without actually obtaining the solution $\mathbf{X}(t)$.

## 4.2 The Liapunov Function

Let us consider the nonlinear autonomous system (4.1) and let $\mathbf{X}=0$ be a point of equilibrium. Then, we define a *Liapunov function* $V(\mathbf{X})$ for the system as follows:

- $V(\mathbf{X})$ and all its partial derivatives $\frac{\partial V}{\partial x_i}$ ($i=1,2,3,...,n$) are continuous, in a nbd. of $\mathbf{X}=0$.
- $V(\mathbf{X})$ is positive definite, i.e. $V(\mathbf{X})>0$, for $\mathbf{X}\neq 0$, in some nbd. $\|\mathbf{X}\|\leq k$ of the origin.
- $\dot{V}$ is *negative semidefinite*, i.e. $\dot{V}(\mathbf{X})\leq 0 \ \forall \ \mathbf{X}$ with $\|\mathbf{X}\|\leq k$ and $\dot{V}(0)=0$, where,

$$\dot{V} = \frac{\partial V}{\partial x_1}\dot{x}_1 + \frac{\partial V}{\partial x_2}\dot{x}_2 + \cdots\cdots + \frac{\partial V}{\partial x_n}\dot{x}_n$$

$$= \frac{\partial V}{\partial x_1}f_1 + \frac{\partial V}{\partial x_2}f_2 + \cdots\cdots + \frac{\partial V}{\partial x_n}f_n$$

**NOTES:**

- A function $V(\mathbf{X})$ continuous in a nbd. of $\mathbf{X}=0$, with $V(0)=0$ is also called a *V-function* of the system (4.1).
- If $V(\mathbf{X})$ is finite as $\|\mathbf{X}\|\to\infty$, then the V-function is called *radially bounded*. On the other hand, if $V(\mathbf{X})\to\infty$ as $\|\mathbf{X}\|\to\infty$, then it is called *radially unbounded*. For example, $V(x_1,x_2)= x_1^2 + \frac{x_1^2}{1+x_1^2}$ is radially bounded. But, $V(x_1,x_2)= x_1^2+x_2^2$ is radially unbounded.
- If $V(\mathbf{X})$ is a radially unbounded positive definite function, then the set of all points $\mathbf{X}$, such that $V(\mathbf{X})=k$ (where, $k>0$) forms a simple closed surface and further the surface $V(\mathbf{X})=k_1$ lies entirely inside the surface $V(\mathbf{X})=k_2$, when $k_1<k_2$.

In the following sections, we shall now see that the nature of the function $V(\mathbf{X})$ determines the stability of the system (4.1).

**Theorem 4.1:** (**Liapunov's stability theorem**)

Consider the nonlinear autonomous system (4.1) and let the origin be an equilibrium point for it. Then, at the origin, the system is

(a) Stable if $\exists$ a Liapunov function $V(\mathbf{X})$ for the system in a region $U = \{x\in R^n \,|\, \|x\|\leq \varepsilon\}$.

(b) Asymptotically stable if $\exists$ a Liapunov function $V(\mathbf{X})$ so that $\dot{V}$ is negative definite in the region $U$.

(c) Globally asymptotically stable if $\exists$ a Liapunov function $V(\mathbf{X})$ so that $\dot{V}$ is negative definite in $U$ and also $V(\mathbf{X})$ is radially unbounded.

**Proof:** Let us prove the theorem for a two dimensional case.

The set $C=\{\mathbf{X}\in R^n \,|\, \|\mathbf{X}\|=\varepsilon\}$ is a closed set. Now as $V(\mathbf{X})$ is continuous on $U$, so clearly it is also continuous on $C$. Thus, $V(\mathbf{X})$ assumes its bounds on $C$. Let, $m$ be the minimum value of $V(\mathbf{X})$ on $C$, i.e. $V(\mathbf{X}) \geq m \ \forall \ \mathbf{X} \in C$. Then, from definition of $V(\mathbf{X})$, we must have $m>0$. Let $C_1$ be the closed set representing $V(\mathbf{X})=m$. Then, $C_1$ is a closed curve.

Now, as $V(\mathbf{X})$ is continuous on $U$, so it is also continuous at $\mathbf{X}=0$ at which $V=0$. Thus, from definition of continuity, for given $r>0$, $\exists\ \delta>0$, such that

$$|V(\mathbf{X}) - V(0)| \leq r, \text{ when } \|\mathbf{X}\| < \delta \tag{4.2}$$

In particular, let $r=k<m$. Then by (4.2) $\exists\ \delta>0$, such that

$$|V(\mathbf{X})| \leq k, \text{ or } V(\mathbf{X}) \leq k\ \forall\ \mathbf{X} \text{ with } \|\mathbf{X}\| < \delta.$$

Let $C_2$ be the closed curve representing $V(\mathbf{X})=k$. Then, $C_2 \subseteq C_1$, i.e. $C_2$ is contained in $C_1$ as shown in Fig. 4.1 below:

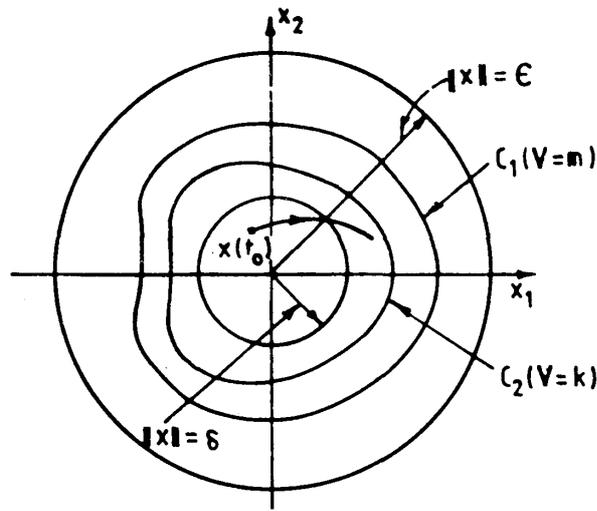

**Fig 4.1: Depiction of the regions, represented by different sets**

Now, obviously $\delta<\varepsilon$, otherwise the assumption of $\min\{V(\mathbf{X}), \|\mathbf{X}\|=\varepsilon\}=m$ will be contradicted.

Let, any $t=t_0$ such that $\|\mathbf{X}(t_0)\| \leq \delta$. Then, as by assumption, $\dot{V} \leq 0$, so we have, $V(\mathbf{X}(t)) \leq V(\mathbf{X}(t_0)) \leq k, \forall\ t>t_0$. So, $V(\mathbf{X}(t))$ can never reach the closed curve $C_1$ and so $\mathbf{X}(t)$ cannot reach the closed set $C$ as $V(\mathbf{X}) \geq m$ on this set.

Thus, we have, $\|\mathbf{X}(t)\| \leq \varepsilon\ \forall\ \|\mathbf{X}(t_0)\| \leq \delta$ with $t>t_0$. So $\mathbf{X}=0$ is a point of *stable* equilibrium.

Now, if $\dot{V}<0$, then as $t$ increases, the curve $C_2$ will shrink towards the origin. So, $\mathbf{X}(t) \to 0$ as $t \to \infty$, i.e. the system (4.1) is *asymptotically stable* at the origin.

Finally, the last condition implies that $C_2=\{\mathbf{X}, V(\mathbf{X})=k\}$ are always closed surfaces in the entire region. Thus, any solution $\mathbf{X}(t)$ starting at $\mathbf{X}=0$ can never go outside the region. This ensures the *globally asymptotical* stability of the system at the origin.

*Hence, the theorem is proved.*

**NOTE: Important remarks about the theorem**
- ➤ The method given by Theorem 4.1 for stability analysis of a nonlinear system is called *Liapunov's Direct Method*.
- ➤ The conditions of this theorem are sufficient but not necessary for stability. If, while applying the theorem, the stability conditions are not satisfied for one V-function, we cannot say that the system is unstable because there may exist another V-function, satisfying the stability requirements.
- ➤ The main problem with this method is that there is no general rule for finding a suitable V-function for a nonlinear system. Also, for a given V-function there is no general rule for checking its positive definiteness.

**Theorem 4.2: (Liapunov's instability theorem)**
Consider the nonlinear autonomous system (4.1) having the origin as an equilibrium point. Let, there is a scalar function $W(\mathbf{X})$, satisfying the following properties:

(a) $W(\mathbf{X})$ is continuous and has continuous partial derivatives $\frac{\partial W}{\partial x_i}$ ($i=1,2,3,...,n$).

(b) $W(\mathbf{X}) \geq 0$, for $\mathbf{X} \neq 0$ and $W(0)=0$.

(c) $\dot{W}$ is positive definite, i.e. $\dot{W}(\mathbf{X}) > 0$ for $\mathbf{X} \neq 0$.

Then, the system is unstable at the origin.

**NOTE:** Again, the conditions of this theorem are sufficient but not necessary for instability of a nonlinear system. Also, like the Liapunov's V-function, there is no general method of finding a suitable W-function.

### 4.3 Domain of Attraction

Let, the origin be an asymptotically stable equilibrium point of the system (4.1). Then, the total set of all initial points $\mathbf{X}$ for which $\mathbf{X}(t) \to 0$ as $t \to \infty$, is called the *Domain of Attraction* for the system. Knowledge of this domain is of great value in practical problems as it enables permissible deviations from equilibrium to be determined.

The region of asymptotic stability obtained by a particular Liapunov's function may be in general only a part of the domain of attraction.

### 4.4 Applications of the Liapunov's Direct Method

Here, we provide a number of mathematical examples to illustrate the applications of the Liapunov's direct method in analyzing stability properties of autonomous systems.

**Example 4.1:** Let us consider the nonlinear system, described by the equations

$$\left. \begin{array}{l} \dot{x}_1 = x_2 \\ \dot{x}_2 = -x_1 - x_2^3 \end{array} \right\}$$

The only equilibrium point of the system is the origin.

Let, $V = x_1^2 + x_2^2$. Then, $V$ is positive definite and has continuous first order partial derivatives with respect to $x_1$ and $x_2$. Now, we have

$$\dot{V} = 2x_1\dot{x}_1 + 2x_2\dot{x}_2$$
$$= 2x_1x_2 + 2x_2(-x_1 - x_2^3)$$
$$= -2x_2^4$$

So, $\dot{V} \leq 0$ for all values of $x_2$. Thus, $\dot{V}$ is negative semidefinite.

So, we have obtained a Liapunov's function $V$ for the given system. Hence, the given nonlinear system is *globally stable* at the origin.

**Example 4.2:** Let us consider the nonlinear system, given by the differential equations

$$\left.\begin{array}{l}\dot{x}_1 = -x_1(1 - 2x_1x_2) \\ \dot{x}_2 = -x_2\end{array}\right\}$$

Here, the only equilibrium point is the origin.

Let, $V = \frac{1}{2}x_1^2 + x_2^2$. Then, $V$ is positive definite and it has first order partial derivatives with respect to $x_1$ and $x_2$. Now, we have

$$\dot{V} = x_1\dot{x}_1 + 2x_2\dot{x}_2$$
$$= -x_1^2(1 - 2x_1x_2) - 2x_2^2$$

Now, $\dot{V} < 0$ if $1 - 2x_1x_2 > 0$, i.e. if $x_1x_2 < \frac{1}{2}$.

Thus, in the region, bounded by all points for which $x_1x_2 < \frac{1}{2}$, the given nonlinear system is asymptotically stable at the origin. However, for this case it is not possible to make a general statement, regarding global stability.

**Example 4.3:** Consider, the nonlinear autonomous system, described by the equations

$$\left.\begin{array}{l}\dot{x}_1 = -2x_1 + x_1x_2 \\ \dot{x}_2 = -x_2 + x_1x_2\end{array}\right\}$$

Here we have two equilibrium points (0,0) and (1,2). We shall study the stability of the origin.

Let, $V = x_1^2 + x_2^2$. Then, $V$ is positive definite and has continuous first order partial derivatives with respect to $x_1$ and $x_2$. Now we have

$$\dot{V} = 2x_1\dot{x}_1 + 2x_2\dot{x}_2$$
$$= 2x_1(-2x_1 + x_1x_2) + 2x_2(-x_2 + x_1x_2)$$
$$= 2x_1^2(x_2 - 2) + 2x_2^2(x_1 - 1)$$

For asymptotic stability, we need $\dot{V}<0$, i.e.

$$x_1^2(x_2-2)+x_2^2(x_1-1)<0 \tag{4.3}$$

All these initial states for which **X**(*t*) (*t*≥0) lies in the region with $\dot{V}<0$ lead to asymptotic stability. The limiting condition for such a region is $\dot{V}=0$, i.e.

$$x_1^2(x_2-2)+x_2^2(x_1-1)=0 \tag{4.4}$$

We can easily verify that the condition, represented by the inequality (4.3) is satisfied in the entire third quadrant. Similarly, information about the other quadrants can be obtained. Also we can see that $V = x_1^2 + x_2^2 = 4$ is the largest region inside which $\dot{V}<0$, i.e. $\dot{V}$ is negative definite and so the system is asymptotically stable. Outside this region the system is possibly unstable. In Fig. 4.2 below, we can see the region of stability of the given system.

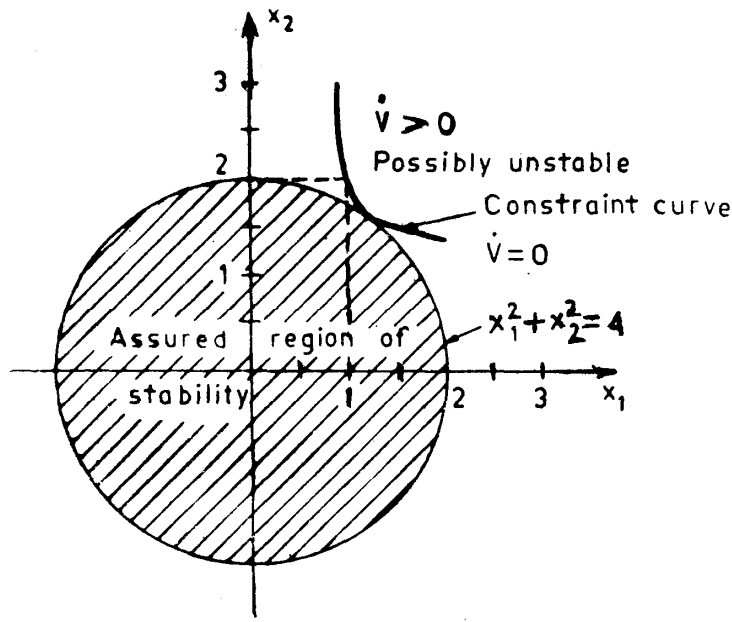

**Fig 4.2: Region of stability of the given autonomous system**

Here, we see that in the entire shaded portion, the system is asymptotically stable at the origin.

**Example 4.4:** Consider a unit mass suspended from a fixed support by a spring as shown in the Fig.4.3. Let us take,

        *z*=displacement from the equilibrium position at any time *t* in the downward direction.
        *k*= spring constant.
        *l*= equilibrium length of the spring.
Let the spring obeys *Hooke's Law* which states that:

*If, the deformation is small, then the stress in a body is proportional to the corresponding strain.*

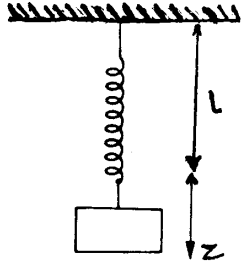

**Fig 4.3: Unit mass suspended by a string**

Now, here at the equilibrium position, the mass is at rest. So, we have $g=kl$ (as unit mass is considered). So, the equation of motion is given by

$$\ddot{z} = -k(l+z) + g$$
$$\Rightarrow \ddot{z} = -kz \qquad [\because g = kl]$$
$$\Rightarrow \ddot{z} + kz = 0$$

i.e.,

$$\ddot{z} = -kz \tag{4.5}$$

This is a nonlinear system. We take $x_1 = z$, $x_2 = \dot{z}$. Then, the system, represented by Eq. (4.5) becomes

$$\left.\begin{array}{l}\dot{x}_1 = x_2 \\ \dot{x}_2 = -k\dot{x}_1\end{array}\right\} \tag{4.6}$$

Clearly the only equilibrium point is the origin. Since, the system is conservative, so the total energy is given by

E= K.E. + P.E.

So, $\quad$ E= $\dfrac{1}{2}kx_1^2 + x_2^2$ $\qquad$ (4.7)

As the system is conservative so the total energy E is clearly positive definite and also has continuous first order partial derivatives with respect to $x_1$ and $x_2$.

Now we have,

$$\dot{E} = kx_1\dot{x}_1 + x_2\dot{x}_2$$
$$\Rightarrow \dot{E} = kx_1\dot{x}_1 + x_2(-kx_1) \quad \text{[Using the state equations (4.6)]}$$
$$\Rightarrow \dot{E} = 0$$

Thus $\dot{E}$ is negative semidefinite and so is a Liapunov function. Hence, by *Liapunov's stability theorem*, the given system is stable at the equilibrium point origin.

**NOTE:** If now we suppose that the force exerted by the spring is some function $k(x_1)$ where $k(0)=0$ and $k(x_1) \neq 0$, for $x_1 \neq 0$, then the system becomes

$$\left. \begin{array}{l} \dot{x}_1 = x_2 \\ \dot{x}_2 = -k \end{array} \right\}$$

In this case, the total energy is given by,

$$E = \frac{1}{2}x_2^2 + \int_0^{x_2} k(\sigma) d\sigma$$

So, $\dot{E} = x_2(-k) + k\dot{x}_1 = 0$

Thus, again by *Liapunov's stability theorem*, the origin is stable for any nonlinear spring constant, satisfying the above condition.

**Example 4.5:** Let us now consider the system in Ex. 4.4 with a damping force $d\dot{z}$ added to it. Then the equation of motion becomes

$$\ddot{z} + d\dot{z} + kz = 0 \qquad (4.8)$$

Now, we suppose that both $d$ and $k$ are constants and for simplicity let $d=1$ and $k=2$. Then, the system (4.8) can be described in state-space form as follows.

$$\left. \begin{array}{l} \dot{x}_1 = x_2 \\ \dot{x}_2 = -2x_1 - x_2 \end{array} \right\} \qquad (4.9)$$

Here again the origin is the only point of equilibrium. The total energy for the system is given by:

$$E = \frac{1}{2}kx_1^2 + x_2^2$$
$$\Rightarrow \dot{E} = 2x_1(x_2) + x_2(-2x_1 - x_2)$$
$$= -x_2^2$$

So, here $\dot{E} \leq 0$ which is negative semidefinite. Thus, the origin is a point of stable equilibrium. Now, we consider the function,

$$V = 7x_1^2 + 2x_1x_2 + 3x_2^2$$
$$\Rightarrow \dot{V} = 14x_1x_2 + 2[x_2(x_2) + x_1(-2x_1 - x_2)] + 6x_2(-2x_1 - x_2)$$
$$= -4(x_1^2 + x_2^2)$$

Clearly $V$ is positive definite with $\dot{V}$ negative. Also, $V$ is radially unbounded. Hence, this new system is globally asymptotically stable at the origin.

**NOTE:** This example illustrates that a suitably chosen Liapunov function can in general provide more information regarding stability than the usual energy function.

**Theorem 4.3:** If in the Liapunov's stability theorem, there exist a Liapunov function $V$, such that $\dot{V}$ does not vanish identically on any non-trivial trajectory of the nonlinear system (4.1), then, the origin is asymptotically stable.

**Example 4.6:** Let us consider the Van der Pol's equation, given by

$$\ddot{E} + \varepsilon(z^2 - 1) + z = 0 \qquad (4.10)$$

Here, $\varepsilon<0$ is a constant. As usual we take $x_1 = z, x_2 = \dot{z}$. Then, the equation (4.10) transforms to

$$\left.\begin{array}{l} \dot{x}_1 = x_2 \\ \dot{x}_2 = -x_1 - \varepsilon(x_1^2 - 1)x_2 \end{array}\right\} \qquad (4.11)$$

Clearly here the origin is the only point of equilibrium. Let us take $V = x_1^2 + x_2^2$ which is clearly positive definite and continuous. Now, we have:

$$\dot{V} = 2x_1\dot{x}_1 + 2x_2\dot{x}_2$$
$$= 2\varepsilon x_2^2(1 - x_1^2)$$

Thus here $\dot{V} \leq 0$ if $x_1^2 < 1$. So by the Theorem 4.3, the system is asymptotically stable at the origin. It means that all the trajectories starting inside the region $\Gamma: x_1^2 + x_2^2$ converge to the origin, as $t \to \infty$ and $\Gamma$ is therefore the *region of asymptotic stability* here.

It is important to note that here the infinite strip $x_1^2 < 1$ is not the region of asymptotic stability, since a trajectory starting outside $\Gamma$ can move outside this strip while continuing in the direction of $V$ circles and hence lead to divergence, as illustrated in Fig.4.4.

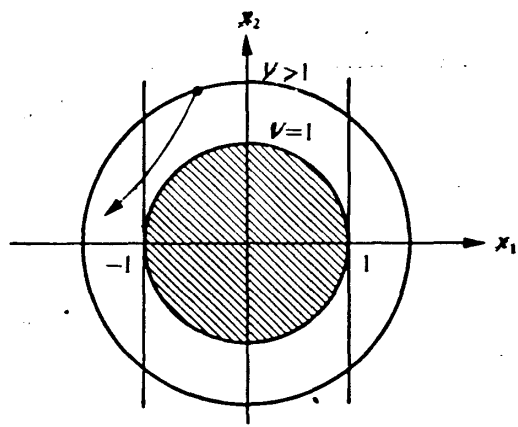

**Fig 4.4: Region of asymptotic stability**

In general if a closed region $R$ is defined by $V(\mathbf{X})=$*constant* is bounded and has $\dot{V}$ negative definite throughout then $R$ is a region of asymptotically stability for the system.

**NOTE:** Suppose that we now take the state variables as

$$x_1 = z, x_3 = \int_0^t z\,dz$$

Then the corresponding state equations are

$$\left.\begin{array}{l} \dot{x}_3 = x_1 \\ \dot{x}_1 = \dot{z} \end{array}\right\}$$

which can be written as

$$\left.\begin{array}{l} \dot{x}_3 = x_1 \\ \dot{x}_1 = -x_3 - \varepsilon(\frac{1}{3}x_1^3 - x_1) \end{array}\right\}$$

Thus again as usual using $V = x_1^2 + x_3^2$, we see that

$$\dot{V} = 2x_1(-x_3 - \frac{1}{3}\varepsilon x_1^3 + \varepsilon x_1) + 2x_3 x_1$$

$$= 2\varepsilon x_1^2 (1 - \frac{1}{3}x_1^2)$$

Thus $\dot{V} < 0$ if $x_1^2 < 3$. So the region of asymptotic stability of the new system with these different set of state variables is $x_1^2 + x_3^2 < 3$ which is larger than before.

### 4.5 Application of Liapunov's Theory to Linear Systems

For a linear system there is a standard approach for finding a Liapunov function. Let us consider the linear autonomous system described by the state equation

$$\dot{\mathbf{X}} = \mathbf{A}\mathbf{X} \tag{4.12}$$

We have seen that the stability nature of the system can be studied directly by computing the eigen values of the matrix $\mathbf{A}$. Here we shall give an alternative approach by using *Liapunov theory*. Let, a Liapunov function for the system is given by $V = x^T P x$ where $P$ is a positive definite symmetric matrix. Now we have,

$$\dot{V} = \dot{x}^T P x + x^T P \dot{x}$$
$$= (Ax)^T P x + x^T P(Ax)$$
$$= x^T (A^T P + PA) x$$

As $\dot{V} \leq 0$, so for the system to be stable we must have $A^T P + PA = -Q$.

Thus, we get for the system (4.12) $\dot{V} = -x^T Qx$. It can be seen that $Q$ is also a symmetric matrix. Now we give the following important theorem for determining the stability of a linear autonomous system.

**Theorem 4.4:** A necessary and sufficient condition for a linear autonomous system, given by (4.12) to be asymptotically stable at the origin is that for any positive definite symmetric matrix Q, there exists a positive definite symmetric matrix P which satisfies the continuous Liapunov matrix equation given by $A^T P + PA = -Q$.

**NOTE:** This theorem does not depend upon the choice of $Q$. So it is convenient to take $Q=I$, the identity matrix.

**Example 4.7:** Let us consider the linear system given by

$$\dot{\mathbf{X}} = \mathbf{A}\mathbf{X} = \begin{bmatrix} 0 & 1 \\ -2 & -3 \end{bmatrix} \mathbf{X}.$$

We want to determine the stability nature of the system at the origin.

Let $Q=I$ and $P = \begin{bmatrix} p_1 & p_2 \\ p_3 & p_4 \end{bmatrix}$. Then, the Liapunov matrix equation gives:

$$A^T P + PA = -I$$

$$\Rightarrow \begin{bmatrix} 0 & -2 \\ 1 & -3 \end{bmatrix} \begin{bmatrix} p_1 & p_2 \\ p_3 & p_4 \end{bmatrix} + \begin{bmatrix} p_1 & p_2 \\ p_3 & p_4 \end{bmatrix} \begin{bmatrix} 0 & 1 \\ -2 & -3 \end{bmatrix} = -I$$

$$\Rightarrow \begin{bmatrix} -4p_2 & p_1 - 3p_2 - 2p_3 \\ p_1 - 3p_2 - 2p_3 & 2p_2 - 6p_3 \end{bmatrix} = \begin{bmatrix} -1 & 0 \\ 0 & -1 \end{bmatrix}$$

Equating corresponding terms from both sides we get,

$$P = \begin{bmatrix} \dfrac{5}{4} & \dfrac{1}{4} \\ \dfrac{1}{4} & \dfrac{1}{4} \end{bmatrix}$$

The matrix $P$ is positive definite and symmetric. Hence by Theorem 4.4, the given system is asymptotically stable at the origin.

*Chapter-5*

# STABILITY OF NON-AUTONOMOUS SYSTEMS

## 5.1 Study of Non-autonomous Systems

So far, we have discussed about the stability properties of linear and nonlinear autonomous systems. Methods for checking the stability properties for such systems have been described. It has been observed that the Liapunov's direct method has a crucial significance in the analysis of stability for an autonomous system. Below, we briefly summarize the key results and observations of the preceding chapters.

- In case of linear autonomous system, we can readily determine the nature of stability by finding the eigen values of the coefficient matrix.
- In the cases where some of the eigen values of the coefficient matrix are zeros, the first method becomes inapplicable and then we use the famous **Routh-Hurwitz** criterion. This criterion covers a large number of autonomous systems.
- One of the well-known methods for studying the local stability of an isolated equilibrium point for an autonomous nonlinear system is by **linearization** of the model, provided it is possible.
- In many cases the information obtained from linearization of a nonlinear autonomous system become inconclusive and so do not clearly reflect the stability nature of the system. In such cases, we use the celebrated **Liapunov's Direct Method.** This method is so far the best known method for stability analysis of both autonomous as well as non-autonomous systems.

Although autonomous or time-invariant systems are relatively easier to analyze with robust techniques for studying their stability properties at our disposal, but most of the practical systems we encounter are of time-variant nature which cannot be studied so easily like their autonomous counterparts. Also the results we have come across in case of autonomous systems are quite inapplicable in case of non-autonomous systems. Now-a-days, the theory of stability of non-autonomous systems is widely used in Physics, Astronomy, Chemistry and even in Biology. It is of primary importance in technology because ships, airplanes and rockets should always keep a prescribed, stable state while moving; turbines and generators should maintain a stable nature; a gyroscopic compass should indicate a stable direction of a geographic meridian and so on.

All these systems stated above contain the associated time component. So to deal with their stability criteria, we must have to consider the non-autonomous case. Thus, we need to develop some additional mathematical theories by which the study of time-variant systems can be performed which otherwise is too difficult.

## 5.2 Some Mathematical Examples of Non-autonomous Systems

In order to motivate our treatment of non-autonomous systems and to have an idea of their natures, we at first begin with examples of some simple systems whose state spaces consist of just a single variable.

**Example 5.1:** Let us consider the very simple system:

$$\frac{dx}{dt} = \sin t \tag{5.1}$$

With the given initial condition $x(0)=x_0$.

First of all, note that this system has no true equilibrium points because here we can see that $\frac{dx}{dt}=0$ when $\sin t=0$. This happens when $t=n\pi$ for any integer $n$. However, since $t$ never stops changing, $dx/dt$ is never zero for more than an instant. Thus $x(t)$ cannot come at a definite equilibrium position at any instance. The state space $x$ is no longer a proper phase space for non-autonomous differential equations because the behavior at a given point in the state space depends on the time at which that point was reached. Essentially, this means that the phase space is two-dimensional and consists of the variables $x$ and $t$. We can in fact formalize this idea by expanding our original equation, represented in Eq. (5.1) into two separate equations:

$$\left. \begin{array}{l} \dfrac{dx}{d\tau} = \sin t \\ \dfrac{dt}{d\tau} = 1 \end{array} \right\} \tag{5.2}$$

In practical terms, we haven't gained much by doing so. Conceptually however this transformation into two equations highlights the fact that a non-autonomous system with a $d$-dimensional state space is in a sense is equivalent to a $d+1$-dimensional autonomous system. The evolution of the time variable is of course trivial. So, a non-autonomous system is not equivalent to a $d+1$-dimensional autonomous system of the general form. Accordingly, **Hale and Kocak** in their book: ***Dynamics and Bifurcations*** call these special non-autonomous systems as "*d and a half-dimensional system*".

The current example is a 1½-dimensional system. There are of course some benefits of such naming, although it is generally preferred to say simply a non-autonomous system. The phase space for the system (5.1) is depicted in Fig. 5.1 below.

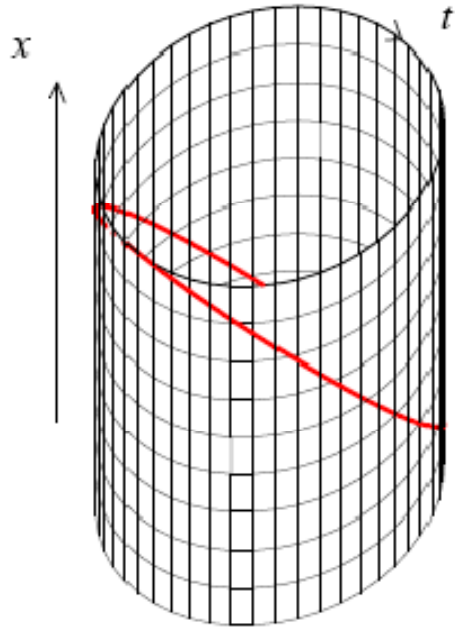

**Fig 5.1: Phase space for the non-autonomous system (5.1)**

Non-autonomous terms don't have to be periodic, but in applications they often are. That being the case, there is some additional structure to the equations we can exploit. The solution of our differential equation (5.1) is: $x(t)=x_0+1-\cos t$. We can draw these solutions in our $(t, x)$ phase space. They are of course just sine waves, the behavior repeats whenever $t$ has increased by $2\pi$, i.e. here the values of the time $t$ are periodic with period $2\pi$. Thus we only need to know the values of time $t$ modulo $2\pi$, i.e. the remainder after subtracting out any integer multiples of $2\pi$ from $t$. It means that we could think of our phase space in this case not as a plane but as a cylinder, where $t$ goes from 0 to $2\pi$ and then repeats again. Varying the initial conditions, we can get different intersections of the curves in space for different values of $x_0$.

**Example 5.2:** Let us consider the differential equation

$$\frac{dx}{dt} = -x + \cos t \qquad (5.3)$$

With the initial condition as $x(0) = 1/2$ at time $t=0$.

Now the system has a unique solution, given by

$$x(t) = \frac{1}{2}(\cos t + \sin t). \qquad (5.4)$$

The equilibrium points of the system are given by:

$$\frac{dx}{dt} = 0$$

$$\Rightarrow \frac{1}{2}(\cos t - \sin t) = 0.$$

$$\Rightarrow \cos t = \sin t$$

$$\Rightarrow t = 2n\pi + \frac{\pi}{4}$$

This is a *limit cycle* in the $x \times (t \bmod 2\pi)$ phase space. By a limit cycle, we roughly mean that the state of a deviated system follows a closed trajectory within the specified tolerance limit and corresponds to an oscillation of fixed amplitude and period. From above it is also clear that the system (5.3) returns to its general equilibrium position if it is slightly displaced from that position. So, the given non-autonomous differential equation is stable at its equilibrium points. But, we can see that the system is not asymptotically stable because $x(t)$ does not approaches to zero, when $t$ increases indefinitely.

### NOTE: An important observation

The two foregoing examples hint that most of the behaviors which are possible in a planar dynamical system are also possible in a non-autonomous system. The one thing we can't have in such a system is a true equilibrium point, unless of course we extend our notion of equilibrium.

**Example 5.3:** Consider the non-autonomous differential equation

$$\frac{dx}{dt} = -x + \frac{1}{t} - \frac{1}{t^2}, \quad t > 0. \tag{5.5}$$

The general solution of the above differential equation is given by:

$$x(t) = \frac{1}{t} + Ce^{-t}, \; t > 0 \text{ where } C \text{ is any arbitrary constant.}$$

Thus this system has an equilibrium point at $(+\infty, 0)$ which is virtually reached for all initial conditions with $t>0$. We notice that the system actually does not attain its equilibrium position in finite time. Also, as $t \to \infty$, $dx/dt \to -x$. It follows that we eventually get solutions which exponentially decay to zero. Since, here the non-autonomous term is not of periodic nature so the phase space is now the entire $(t, x)$ plane.

In some cases it is quite easy to work directly with the equilibrium points obtained at infinity as we have done here. But sometimes we transform the given system to a new system so as to get the points of equilibria at finite coordinates. Let us transform our system into a planar system as below.

$$\left. \begin{array}{l} \dfrac{dx}{d\tau} = -x + \dfrac{1}{t} - \dfrac{1}{t^2}, \; t > 0 \\[2mm] \dfrac{dt}{d\tau} = 1 \end{array} \right\} \tag{5.6}$$

Now we want to change our coordinate system such that the interval [0, ∞) is mapped into some finite interval, say [0, 1]. There are many possible transformations, of which we use the following one:

$$\theta = \frac{t}{t+1}$$

By chain rule, we have from the above transformation,

$$\frac{d\theta}{d\tau} = \frac{d\theta}{dt}\frac{dt}{d\tau}$$

$$= \frac{1}{(t+1)^2}$$

Also, we have $t = \frac{\theta}{1-\theta}$. Thus,

$$\left.\begin{array}{l} \dfrac{d\theta}{d\tau} = (1-\theta)^2 \\ \dfrac{dx}{d\tau} = -x + \dfrac{1-\theta}{\theta} - (\dfrac{1-\theta}{\theta})^2 \end{array}\right\} \quad (5.7)$$

Now the point $(\theta^*, x^*) = (1,0)$ is clearly an equilibrium point of the transformed system (5.7). The system is now in a form for which we can study the stability at the point (1,0) and hence the stability of the equilibrium point at infinity of the original non-autonomous system.

**Example 5.4:** Consider the following second-order *mass-spring damper* of unit mass system with time-varying damping:

$$\ddot{x}(t) + c(t)\dot{x}(t) + k_0 x(t) = 0, \quad x(0) = x_{10}, \quad \dot{x}(0) = x_{20} \quad (5.8)$$

where $c(t) \geq 0$ is the time-varying damping coefficient and $k_0$ is the spring constant. Of course one can think that as long as $c(t)$ is strictly greater than some positive constant, the system should asymptotically return to its original equilibrium point which is $(x, \dot{x}) = (0,0)$. But this is the case with the autonomous second-order *mass-spring damper* system with *constant damping* which is given by the equation:

$$\ddot{x}(t) + c_0\dot{x}(t) + k_0 x(t) = 0, \quad x(0) = x_{10}, \quad \dot{x}(0) = x_{20}, c_0 \geq 0 \quad (5.9)$$

However, if we select $c(t) = 2+e^t$ and $k_0=1$, then the solution of the system (5.8) with the initial conditions $x(0) = 2$, $\dot{x}(0) = -1$ can be written as: $x(t) = 1+e^{-t}$, which tends to $(x, \dot{x}) = (1,0)$ as $t \to \infty$, instead of reaching the equilibrium!! This is a peculiar situation. The conclusion we can draw from this is that the time-varying damping grows so fast that the system gets stuck at $x=1$. We shall discuss about such peculiarities later after taking a close mathematical view of the non-autonomous systems.

**Example 5.5:** We turn to the following linear time-varying equations

$$\dot{x}(t) = A(t)x(t) \tag{5.10}$$

Here the situation is much more complicated. It might be thought that as in case of linear time-invariant systems, if the real parts of all the eigen values of the matrix A(t) are negative for all $t \geq t_0$, then the system would be asymptotically stable at the origin. Unfortunately this conjecture is not true in this case. It can be seen through the example below.

Let us consider the *decoupled system*, given by the equations:

$$\left. \begin{array}{l} \dot{x}_1 = -x_1 + x_2 e^{2t} \\ \dot{x}_2 = -x_2 \end{array} \right\} \text{ which can be written as:}$$

$$\begin{bmatrix} \dot{x}_1 \\ \dot{x}_2 \end{bmatrix} = \begin{bmatrix} -1 & e^{2t} \\ 0 & -1 \end{bmatrix} \begin{bmatrix} x_1 \\ x_2 \end{bmatrix}, \; x_1(0) = x_{10}, \; x_2(0) = x_{20} \tag{5.11}$$

Here the coefficient matrix is $A(t) = \begin{bmatrix} -1 & e^{2t} \\ 0 & -1 \end{bmatrix}$.

This matrix A(t) has all negative eigen values -1, -1, for all $t \geq 0$. Yet solving the system (5.11), we can get:

$$x_2(t) = x_2(0)e^{-t}, \; \dot{x}_1(t) + x_1(t) = x_2(0)e^{t} \tag{5.12}$$

This solution basically demonstrate that the component $x_1(t)$ can escape to infinity as $t$ becomes large enough, since, it contains the unbounded input term $e^t$ and $e^t \to \infty$ as $t \to \infty$.

**Example 5.6:** Consider the following simple non-autonomous system:

$$x(t) = t \tag{5.13}$$

Now here $\dfrac{dx}{dt} = 1$. Thus in this case $\dot{x}(t)$ can never be zero at any instance. So the system has no equilibrium point. This means that there also exist non-autonomous systems which lack equilibria. Let us see the above system in a 2D plane in Fig. 5.2 below.

- 42 -

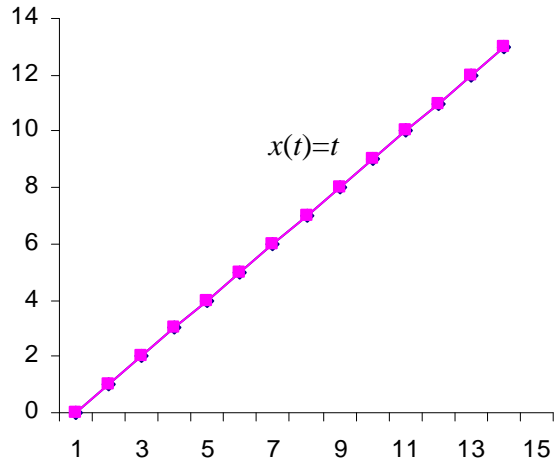

**Fig 5.2: 2D plane diagram of the system (5.13)**

From the above figure it is clear that if a particle slides from any position on the line *x(t)=t* then it cannot rest at equilibrium anywhere.

Although, theoretically systems with no equilibrium points exist in practice (as seen in the above example) but these are of little practical use.

**Example 5.7:** As a final example let us consider the following system:

$$t\dot{x}(t) - x + t = 0, \, t > 0, x(1) = 0 \tag{5.14}$$

This can be written as $\dfrac{dx}{dt} - \dfrac{x}{t} = t, t > 0, x(1) = 0$. Its solution is given by the equation $x(t) = t^2 - t, t > 0.$ So we see that the equilibrium point of the system is given by:

$$\dfrac{dx}{dt} = 0$$
$$\Rightarrow 2t - 1 = 0$$

So, $t = 1/2$. But, according to the given conditions *t*>0 and the initial time is *t*=1. The equilibrium point is $x(0.5) = -0.25$ at $t = 0.5$, which lies in the time interval $[0,1]$. So, although the equilibrium point exists but the system cannot actually attain the equilibrium in this case. Now, we visualize the above situation diagrammatically as below.

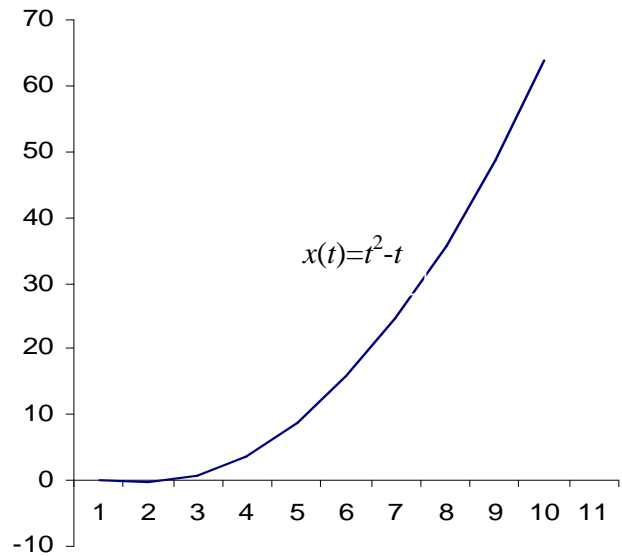

**Fig 5.3: 2D plane diagram of the system (5.14)**

From Fig. 5.3, it is clear that a particle sliding down from any position on the curve cannot come to rest under the given initial conditions. Thus although at time $t=0.5$ mathematically the equilibrium point exists, but practically the above system has no equilibrium point at all.

### NOTE: An important remark

Thus we see that the existence of the equilibrium points of a time-varying system also depends upon the initial conditions prescribed. For practical systems it can be possible that a system attains equilibrium in case of one set of suitable initial conditions but not in case of others.

### NOTE: Conclusions drawn

These examples illustrates that the non-autonomous systems are not so well behaved like the autonomous systems. Various peculiarities are often encountered while dealing with these systems. So, without performing a careful study it is not wise to jump into any conclusion about the nature of stability of such a system.

Now we are going to mathematically define the stability of a non-autonomous system.

### 5.3 Mathematical Definition of Stability of Non-autonomous Systems

We have already defined the stability of an autonomous system in the sense of Liapunov. Non-autonomous systems present a more general class of systems and therefore we need to generalize the concepts of equilibrium and definitions of stability for these systems.

Let us consider the general non-autonomous system, given by:

$$\dot{x}(t) = f(x(t), t), \ x(t_0) = x_0, \ x \in R^n \qquad (5.15)$$

Here $f:D\times[t_0,\infty)\to R^n$ is piecewise continuous in $t$ and locally Lipschitz in $x$ on $D\times[t_0,\infty)$, where $D$ is an open set containing the origin $x=0$. Also we assume that the initial time $t_0\geq 0$. Without loss of generalization we can take $t_0=0$, so that in this case we have the function as $f:D\times R^+ \to R^n$. The equilibrium point $x^*$ of the system (5.15) is defined as the solution of the following system of equations: $f(x^*,t)\equiv 0\ \forall\ t\geq t_0$. This implies that if once the system state is at $x^*$, it remains there for all $t\geq t_0$.

By a suitable transformation, we can make the origin $x=0$ as the equilibrium point of the above system. Taking origin as the equilibrium point of the system we now define the following.

The equilibrium point $x=0$ of the system (5.15) is said to be

- *Stable* at $t_0$ if for any $\varepsilon>0$, there exists a real number $\delta=\delta(\varepsilon,t_0)>0$, such that $\|x(0)\|\leq\delta \Rightarrow \|x(t)\|\leq\varepsilon,\ \forall\ t\geq t_0>0$.
- *Convergent* at $t_0$ if there exists a real number $\delta_1 = \delta_1(t_0)>0$, such that $\|x(0)\|\leq\delta_1 \Rightarrow \lim_{t\to\infty} x(t)=0$. Equivalently (and more precisely), $x_0$ is convergent at $t_0$ if for any given real number $\varepsilon_1>0$, there exists a real number $T=T(\varepsilon_1,t_0)$, such that $\|x(0)\|\leq\delta_1 \Rightarrow \|x(t)\|\leq\varepsilon_1\ \forall\ t\geq t_0+T$.
- *Asymptotically stable* at time $t_0$ if it is both stable and convergent at $t_0$.
- *Unstable* at $t_0$ if it is not stable at $t_0$, i.e. if there is at least one $\varepsilon>0$, so that $\forall\ \delta>0$, $\|x(0)\|<\delta \Rightarrow \|x(t_1)\|>\varepsilon$ for some $t_1\geq t_0>0$.

During the above definitions, we have considered the time component $t_0$. If this dependence is removed we can get the notions of uniform stability which is very much expected for any practical system. Let us now define this concept.

The equilibrium point $x=0$ of the system (5.15) is said to be

- *Uniformly stable* if for any given $\varepsilon>0$, there exists $\delta=\delta(\varepsilon)>0$, such that $\|x(0)\|\leq\delta \Rightarrow \|x(t)\|\leq\varepsilon\ \forall\ t\geq t_0>0$. Here the number $\delta$ depends only on $\varepsilon$ and is independent of the time component $t_0$.
- *Uniformly convergent* if there is $\delta_1 > 0$, independent of the time $t_0$, such that $x_0\leq\delta_1 \Rightarrow x(t)\to 0$ as $t\to\infty$. Equivalently, $x = 0$ is uniformly convergent if for any given real number $\varepsilon_1>0$, there exists a number $T=T(\varepsilon_1)$ such that $\|x(0)\|\leq\delta_1 \Rightarrow \|x(t)\|\leq\varepsilon_1\ \forall\ t\geq t_0+T$.
- *Uniformly asymptotically stable* if it is uniformly stable as well as uniformly convergent.
- *Globally uniformly asymptotically stable* if it is uniformly asymptotically stable and every motion converges to the origin. Here $\delta$ and $\varepsilon$ can be chosen arbitrarily large.

In the above definitions, norm always means the standard Euclidean norm.

**NOTE: Some important special cases**

We have given the general definitions of stability and convergence in case of non-autonomous systems. Now let us see some mathematical implementations of these definitions:

➢ The time-varying system $\dot{x}(t) = A(t)x(t)$ has a unique equilibrium point at the origin $x=0$, provided $A(t)$ is non-singular for all $t \geq 0$.

➢ The nonlinear time-varying system $\dot{x}(t) = x^2(t) + b(t)$ with $b(t) \neq 0$ has no equilibria at all.

➢ The nonlinear non-autonomous system $\dot{x}(t) = (x(t) - 1)^2 b(t)$ with $b(t) \neq 0$ has a unique equilibrium at the point $x=1$.

➢ The nonlinear system $\dot{x}(t) = (x(t) - 1)(x(t) - 2)b(t)$ with $b(t) \neq 0$ has two equilibrium points at $x=1$ and $x=2$.

**NOTE: A remark**

We have just seen that the definitions of stability, asymptotic stability etc. can be extended to time-varying systems with the consideration of the initial time constant $t_0$. An important concept that needs particular attention in the analysis of time-varying systems is *uniformity*. Uniformity in time is important in order to ensure that the region of attraction does not vanish (or tends to zero) with time.

**Example 5.8:** The first order system $\dot{x}(t) = \dfrac{-x}{1+t}$, $x(t_0) = x_0$ has the general solution $x(t) = \dfrac{1+t_0}{1+t} x_0$. So, $x(t)$ converges to zero asymptotically as the time $t \to \infty$, but not uniformly, because larger $t_0$ in the numerator requires longer time for convergence from the same initial condition $x_0$. To get the same time for convergence one needs to decrease $\|x_0\|$ uniformly with the increase of $t_0$, which implies shrinking of the domain of attraction with time.

Thus we have seen that a non-autonomous system may be asymptotically stable though not uniformly stable. Of course uniform asymptotic stability is practically more desirable as it also ensures convergence.

**Example 5.9:** Let us consider the time-varying system:

$$\left.\begin{array}{l} \dot{x}_1(t) = \dfrac{2x_1}{t} + 1 \\ \dot{x}_2(t) = 2t \end{array}\right\}, \text{ where } \left[\begin{array}{c} x_1 \\ x_2 \end{array}\right]_{(t=1)} = \left[\begin{array}{c} 1 \\ 1 \end{array}\right], t > 0 \qquad (5.16)$$

The general solution of the system is given by:

$$\left.\begin{array}{l} x_1(t) = 2t^2 - t \\ x_2(t) = t^2 \end{array}\right\}$$

So, the given system can also be written as follows:

$$\left.\begin{array}{l}\dot{x}_1(t) = \dfrac{2x_1}{t} + 1 \\ \dot{x}_2(t) = 2\sqrt{x_2}\end{array}\right\}, t > 0 \tag{5.17}$$

Here the initial time is $t_0=1$. Now from (5.17) we see that $\dot{x}=0$ for $(\dfrac{-t}{2},0)$ whatever the time $t$ may be. Now, we have:

$$x_1(t) = -\dfrac{t}{2}$$
$$\Rightarrow 2t^2 - t = -\dfrac{t}{2}$$
$$\Rightarrow 4t^2 - t = 0$$
$$\Rightarrow t = 0, 0.25$$

Now, as for equilibrium $\dot{x}_2(t)$ should be also zero and so the only possible value of $t$ is $t=0$. Hence, the only equilibrium point is $x=(0,0)$ at time $t=0$.

From definition it follows that the system is stable at the equilibrium point origin, nut not asymptotically because $x(t)$ does not approach to zero as $t \to \infty$.

**Example 5.10:** A linear time-varying system is given by:

$$\begin{bmatrix}\dot{x}_1 \\ \dot{x}_2\end{bmatrix} = \begin{bmatrix}\dfrac{1}{t} & t \\ 1 & t\end{bmatrix}\begin{bmatrix}x_1 \\ x_2\end{bmatrix}, t > 0, x(0) = x_0 \tag{5.18}$$

It has a unique equilibrium point (0,0) if the coefficient matrix is non-singular which demands that the time $t \neq 1$. Thus, in this case, the system attains an equilibrium iff the initial time $t_0$ is chosen to be greater than unity.

# EXPONENTIAL STABILITY OF NON-AUTONOMOUS SYSTEMS

## 6.1 Concept of Exponential Stability

Very often in realistic applications of asymptotic stability it is not sufficient to know that trajectories of a system converge to the origin when time becomes arbitrarily large; we also need to know in addition, the rate at which they converge. This brings up the important notion of *exponential stability*.

Let us consider the general time-varying system, given by:

$$\dot{x}(t) = f(x(t),t),\ x(t_0) = x_0,\ x \in R^n \tag{6.1}$$

Here, the function *f(t)* as usual satisfies the standard conditions.
The equilibrium point *x*=0 of the system (6.1) is said to be a

- ➢ *Locally exponentially stable (LES)* equilibrium point if there exist two strictly positive numbers $\alpha$ and $\lambda$ such that for some $\delta > 0$

$$\|x(0)\| \leq \delta \Rightarrow \|x(t)\| \leq \alpha \|x(0)\| e^{-\lambda t}\ \forall t \geq t_0 > 0.$$

- ➢ *Uniformly exponentially stable* if there exist two positive numbers $\alpha$ and $\lambda$ such that for sufficiently small $x_0 = x(t_0)$,

$$\|x(t)\| \leq \alpha \|x_0\|\ e^{(-\lambda(t-t_0))}\ \forall t \geq t_0 > 0.$$

- ➢ *Globally (uniformly) exponentially stable* if the definition of exponential stability holds for any arbitrary $x_0$ and $\delta \in [0,\infty)$.

The positive number $\lambda>0$ is called *rate of exponential convergence*. The definition implies that the state of the system, which is exponentially stable, converges to the origin faster than an exponential function. The concept of exponential stability is illustrated through Figs. 6.1 (a), (b) which are drawn over the time axis. It is important to note that exponential stability always implies asymptotic stability, but the opposite is not true. Another thing is that a system to be exponentially stable must have to be uniformly stable first. It is meaningless to say that a system is uniformly unstable but is still exponentially stable.

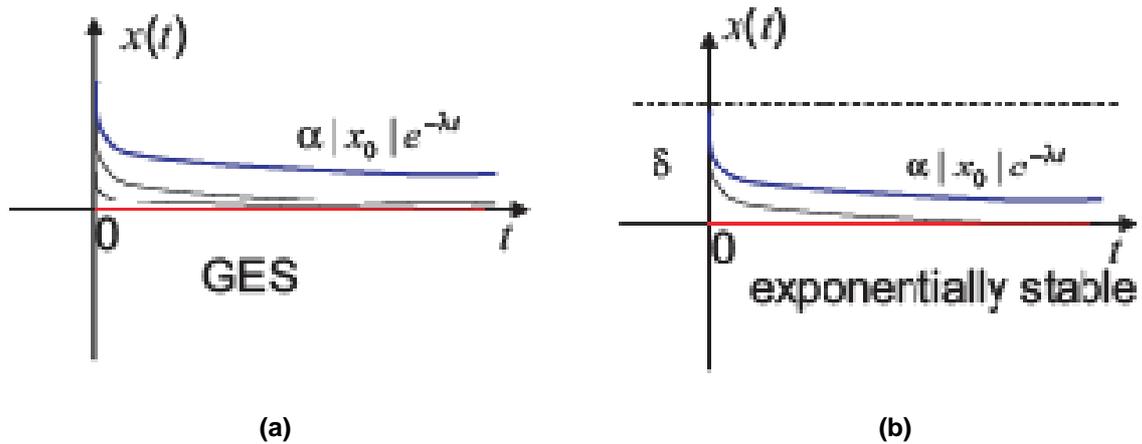

**Fig 6.1:** Depiction of exponential stability:
(a) exponential stability, (b) globally exponential stability

## 6.2 Some Mathematical Examples

Let us see some important mathematical examples of exponential stability of non-autonomous dynamical systems.

**Example 6.1:** Let us consider the system, given by

$$\dot{x}(t) = -(1+\sin^2 x(t))x(t), \quad x(t_0) = x(0) \tag{6.2}$$

The solution of the system is

$$x(t) = x(0)\exp\left(-\int_0^t (1+\sin^2 x(\tau))d\tau\right)$$

$$\Rightarrow x(t) = x(0)e^{\left(\frac{\sin 2t}{4} - \frac{3t}{2}\right)}$$

$$\Rightarrow |x(t)| \leq |x(0)|e^{\frac{1}{4}}e^{-\frac{3t}{2}}$$

$$\Rightarrow |x(t)| \leq e^{\frac{1}{4}}|x(0)|e^{-\frac{3t}{2}}$$

From this solution we can see that $|x(t)| \leq \alpha|x(0)|e^{-\lambda t}$ where $\alpha = e^{\frac{1}{4}}, \lambda = 1.5$. Thus, the trajectories of the system (6.2) exponentially converge to the origin with a rate of convergence $\lambda=1.5$.

**Example 6.2:** We now consider the system given by

$$\dot{x}(t) = -x^2, \quad x(0) = 1 \tag{6.3}$$

This system has the unique solution, given by $x(t) = \dfrac{1}{1+t}$.

Now $x(t) \to 0$ as $t \to \infty$. Thus, $x(t)$ converges to zero slower than any exponential function $e^{-\lambda t}$ with $\lambda>0$. This is because from the solution of the system, we find that $x(t)$ can never approach to zero in a finite time. Thus, in this case the origin is an asymptotically stable equilibrium point but not exponentially stable.

**Example 6.3:** Let us consider the non-autonomous system given by

$$t\dot{x}(t) + x(t) + 2te^{-2t} = 0 \tag{6.4}$$

Here we assume that $x(t)$ is finite and very small when time is large, i.e. $x(t) \to 0$ as $t \to \infty$. Now solution of the above system is found as

$$x(t) = \frac{1}{t}\left(e^{-2t} + 1\right), t > 1 \tag{6.5}$$

From this solution we see that

$$|x(t)| = \left|\frac{1}{t}\left(e^{-2t} + 1\right)\right|$$

$$\Rightarrow |x(t)| = \frac{\left|e^{-2t} + 1\right|}{|t|}$$

$$\Rightarrow |x(t)| \leq \frac{\left|e^{-2t}\right|}{|t|}$$

$$\Rightarrow |x(t)| \leq e^{-2t}$$

The equilibrium point of the system is attained at infinite time and the origin is the only point of equilibrium in this case. From the solution it is clear that the above system is globally exponentially stable with a rate $\lambda=2$. The system is also asymptotically stable at origin.

The graph of the function $x(t)$ is shown in Fig. 6.2. From this phase plane diagram it is clear that the system exponentially advances towards the origin and reaches it at infinite time. It is important to note that the equilibrium point $x=0$ is not practically reached here in any finite time. Thus in this case the system has no finite time equilibrium point.

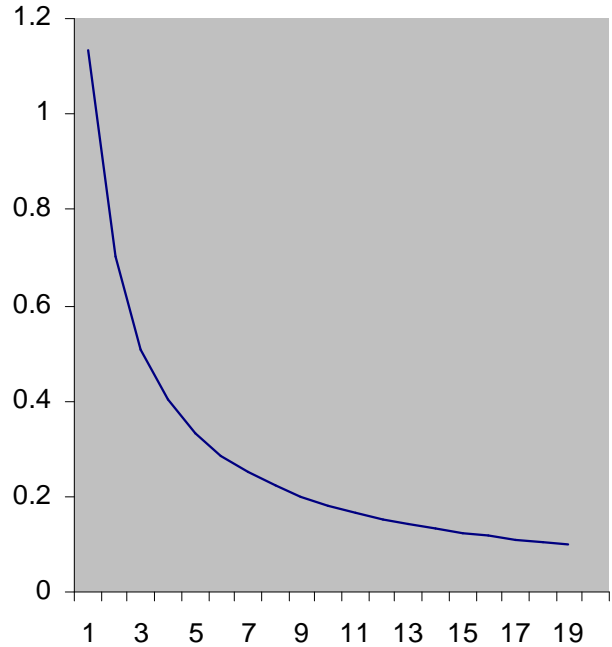

**Fig 6.2: Phase plane diagram of the system (6.4)**

## 6.3 Concept of α-stability

An integral part of the stability analysis of differential equations is the existence of inherent time delays. Time delays are frequently encountered in many physical and chemical processes as well as in the models of hereditary systems, such as the Lotka-Volterra systems, control of epidemics, etc. Therefore, the stability analysis of time-delay systems has received considerable extent of attention during recent years. One of the extended stability properties is the concept of the α-stability which relates to the exponential stability with a convergent rate α>0.

### 6.3.1 α-stability of Retardated Systems

Let us consider the following retardated system, given by

$$\left.\begin{array}{l}\dot{x}(t) = f(t, x(t), x(t-h)), \ t \geq 0 \\ x(t) = \phi(t), \ t \in [-h, 0]\end{array}\right\} \qquad (6.6)$$

The system, represented through Eqs. (6.6) is said to be α-stable with α>0 if there exists a function $\zeta(.)$ such that for each $\phi(.)$, the solution $x(t, \phi)$ of this system satisfies

$$\|x(t, \phi)\| \leq \zeta(\|\phi\|) e^{-\alpha t} \ \forall \, t \geq 0.$$

Here $\|\phi\| = \max\{\|\phi(t)\| : t \in [-h, 0]\}$. This implies that for α>0, the system can be made exponentially stable with the convergent rate α.

### 6.3.2 *α*-stability of Linear Non-autonomous Systems

Let us consider the following linear non-autonomous system with multiple delays given by

$$\left.\begin{array}{l} \dot{x}(t) = A_0(t)x(t) + \sum_{i=1}^{m} A_i(t)x(t-h_i),\ t \geq 0 \\ x(t) = \phi(t),\ t \in [-h, 0] \end{array}\right\} \quad (6.7)$$

where, we have $h = \max\{h_i : i = 1, 2, 3, \ldots, m\}$ and $A_i(t)\,(i = 0, 1, 2, 3, \ldots, m)$ are given matrix functions and $\phi(t) \in C\big([-h, 0], R^n\big)$.

The system (6.7) is said to be *α*-stable with *α*>0 if there exists a function $\zeta(.) : R^+ \to R^+$ such that for each $\phi(t) \in C\big([-h, 0], R^n\big)$ the solution $x(t, \phi)$ of this system satisfies

$$\|x(t, \phi)\| \leq \zeta(\|\phi\|)e^{-\alpha t},\ \forall\ t \in R^+.$$

### 6.3.3 Some Important Notations and Definitions

Here we shall introduce some important notations and definitions which will be frequently used in the foregoing discussions.

- $R^n$ denotes the vector space with scalar product $\langle .,. \rangle$ and vector norm $\|.\|$; also, $R^{n \times r}$ denotes the space of all matrices of dimension $(n \times r)$.
- $\lambda(A)$ denotes the set of all eigen values of the matrix $A$. Also we define the maximum eigen value of $A$ as $\lambda_{\max}(A) = \max\{\text{Re}(\lambda) : \lambda \in \lambda(A)\}$.
- By $\|A\|$ we denote the spectral norm of the matrix $A$ which is defined as

    $$\|A\| = \sqrt{\lambda_{\max}(A^T A)}.$$

- $\eta(A)$ denotes the *matrix measure* of the matrix $A$ which is given by

    $$\eta(A) = \frac{1}{2}\lambda_{\max}(A + A^T).$$

- $C\big([a, b], R^n\big)$ denotes the set of all $R^n$ valued continuous functions on the closed interval $[a, b]$.
- Matrix $A$ is called *positive semidefinite* ($A \geq 0$) if $\langle Ax, x \rangle \geq 0,\ \forall x \in R^n$; also $A$ is called *positive definite* ($A > 0$) if $\langle Ax, x \rangle > 0,\ \forall x \in R^n$.

**Theorem 6.1:** The linear non-autonomous system (6.7) with multiple time delays is $\alpha$-stable if there exists a symmetric positive semidefinite matrix $P(t)$, where $t \in R^+$ such that

$$\left. \begin{array}{l} \dot{P}(t) + [A^T{}_{0,\alpha}(t) + A_{0,\alpha}(t)][P(t) + I] \\ \quad + \sum_{i=1}^{m}[P(t)+I]A_{i,\alpha}(t)A^T{}_{i,\alpha}(t)[P(t)+I] + mI = 0. \end{array} \right\} \quad (6.8)$$

Here $A_{0,\alpha}(t) = A_0(t) + \alpha I$, $A_{i,\alpha}(t) = e^{\alpha h_i} A_i(t), (i=1,2,\ldots,m)$. The equation (6.8) is called the **Riccati Differential Equation (RDE)**.

### NOTE: An important remark

Note that the existence of a positive semidefinite matrix solution $P(t)$ also guarantees the boundedness of the solution and hence the exponential stability of the linear non-autonomous delay system (6.7). Also, while proving the theorem, the stability of the matrix $A(t)$ is not assumed.

**Theorem 6.2:** The linear multiple autonomous time delay system (6.7), where $A_i$ are constant matrices, is $\alpha$-stable if there exists a symmetric positive semidefinite matrix $P \in R^{n \times n}$ which is a solution of the algebraic RDE, given by

$$[A^T{}_{0,\alpha} + A_{0,\alpha}][P+I] + \sum_{i=1}^{m}[P+I]A_{i,\alpha}A^T{}_{i,\alpha}[P+I] + mI = 0 \quad (6.9)$$

**Example 6.4:** Consider the linear non-autonomous time delay system in $R^2$:

$$\dot{x}(t) = A_0(t)x + A_1(t)x(t-0.5) + A_2(t)x(t-1) \; , \; t \in R^+$$

with any given initial function $\phi(t) \in C\left([-1,0], R^2\right)$ and

$$A_0(t) = \begin{bmatrix} a_0(t) & 0 \\ 0 & -7.5 \end{bmatrix}, \; A_1(t) = \begin{bmatrix} e^{-0.5}a_1(t) & 0 \\ 0 & e^{-0.5}\sqrt{3} \end{bmatrix}, \; A_2(t) = \begin{bmatrix} e^{-1}a_1(t) & 0 \\ 0 & e^{-1}\sqrt{3} \end{bmatrix}.$$

We take $a_0(t) = \dfrac{7e^{-9t} - 5}{2(1+e^{-9t})}$, $a_1(t) = \dfrac{1}{\sqrt{2}(1+e^{-9t})}$.

Now, here we have $m=2$, $h_1=0.5$, $h_2=1$. The matrix $A_0(t)$ is not asymptotically stable because we can see that Re($\lambda(A_0)$)=0.5>0. Now let us take $\alpha=1$. Then we have,

$$A_{0,\alpha}(t) = \begin{bmatrix} a_0(t)+1 & 0 \\ 0 & -6.5 \end{bmatrix}, A_{1,\alpha}(t) = A_{2,\alpha}(t) = \begin{bmatrix} a_1(t) & 0 \\ 0 & \sqrt{3} \end{bmatrix}.$$

Then, the solution RDE (6.8) can be found as

$$P(t) = \begin{bmatrix} e^{-9t} & 0 \\ 0 & 1 \end{bmatrix} \geq 0 \; \forall t \in R^+.$$

Here, we see that the matrix $P(t)$ is symmetric, positive semidefinite and satisfies the RDE. Thus all the conditions of the Theorem 6.1 are satisfied and hence the linear non-autonomous delay system is *1-stable* in this case.

**Example 6.5:** Consider the linear autonomous delay system

$$\dot{x}(t) = A_0 x(t) + A_1 x(t-2) + A_2 x(t-4), \; t \in R^+$$

with any given initial function $\phi(t) \in C\left([-4,0], R^2\right)$ and

$$A_0 = \begin{bmatrix} -\frac{17}{6} & 0 \\ \frac{4}{3} & -3.5 \end{bmatrix}, \; A_1 = \begin{bmatrix} e^{-1} & 0 \\ 0 & e^{-1} \end{bmatrix}, \; A_2 = \begin{bmatrix} e^{-2} & 0 \\ 0 & e^{-2} \end{bmatrix}.$$

Here the coefficient matrices are independent of the time $t$.
In this case, $m=2$, $h_1=2$, $h_2=4$. Taking $\alpha=0.5$, we have

$$A_{0,\alpha}(t) = \begin{bmatrix} -\frac{7}{3} & 0 \\ \frac{4}{3} & -3 \end{bmatrix}, \; A_{1,\alpha}(t) = A_{2,\alpha}(t) = \begin{bmatrix} 1 & 0 \\ 0 & 1 \end{bmatrix} = I_2.$$

Now, the solution of the RDE (6.9) can be found as $P = \begin{bmatrix} 1 & -1 \\ -1 & 1 \end{bmatrix}$. Here, the principal minors of the matrix $A$ are, 1, 0. Thus, $P$ is positive semidefinite and also it is symmetric. Hence, the system in this case is *0.5-stable*.

**Theorem 6.3:** Let us consider the multiple time delay system (6.7). Let, the matrix functions $A_i(t)$, $i = 0, 1, 2, \ldots, m$ are bounded on $R^+$ and in addition the following conditions are satisfied:

> $\eta(A_0) = \sup_{t \in R^+} \eta(A_0(t)) < +\infty$ and $P(t)$, $A_i(t)$ are bounded on $R^+$.

> $\eta(A_0) + \alpha \|P_I\| + \frac{m}{2} e^{2\alpha h} \|P_I\|^2 \|A\|^2 \leq 0$                                                             (6.10)

where, we define $\|P_I\| = \sup_{t \in \Re^+}(\|P(t) + I\|)$, $\|A\|^2 = \sup_{t \in \Re^+} \|A(t)\|^2$.

Then, the non-autonomous delay system (6.7) is $\alpha$-stable if the **Liapunov Equation** $\dot{P}(t) + [A^T_0(t) + A_0(t)]P(t) + mI = 0$ has a solution $P(t) \geq 0$, which is bounded on $R^+$. In this case, the rate of convergence $\alpha > 0$ is the solution of the inequality (6.10).

## NOTE: Corollary of the Theorem

The linear time delay system (6.7), where $A_i(t)$ are constant matrices is $\alpha$-stable if there is a symmetric positive semi-definite matrix $P$ satisfying the algebraic Liapunov equation $A_0^T P + PA_0 + mI = 0$. In this case, the convergence rate $\alpha > 0$ is the solution of the following inequality:

$$\eta(A_0) + \alpha \|P_I\| + \frac{m}{2} e^{2\alpha h} \|P_I\|^2 \|A\|^2 \leq 0$$

where, $P_I = P + I$, $\|A\|^2 = \max\{\|A_i\|^2, i = 1, 2, \ldots, m\}$.

**Remark:** Theorem 6.3 is very important in determining the rate of stability of time delay systems. As such, this theorem has various applications in practical problems.

Let us now see some mathematical examples which use Theorem 6.3.

**Example 6.6:** Consider the linear non-autonomous multiple delay system

$$\dot{x}(t) = A_0(t)x + A_1(t)x(t-0.5) + A_2(t)x(t-1) \ , \ t \in R^+$$

with any given initial function $\phi(t) \in C\left([-1,0], R^2\right)$ and

$$A_0(t) = \begin{bmatrix} 0.5 - e^t & 1 \\ -1 & 0.5 - e^t \end{bmatrix}, \ A_1(t) = e^{-0.2} \sin t \begin{bmatrix} \frac{1}{40} & 0 \\ 0 & \frac{1}{40} \end{bmatrix}, \ A_2(t) = e^{-0.2} \cos t \begin{bmatrix} \frac{1}{40} & 0 \\ 0 & \frac{1}{40} \end{bmatrix}.$$

Here, $m=2$, $h_1=0.5$, $h_2=1$. Now, $A_0(t) + A^T{}_0(t) = \begin{bmatrix} 1 - 2e^t & 0 \\ 0 & 1 - 2e^t \end{bmatrix}$. The eigen values of $A_0(t) + A^T{}_0(t)$ are $(1-2e^t)$, where $t \in R^+$. Thus from definition, we have here

$$\eta(A_0) = \sup_{t \in \Re^+} \eta(A_0(t))$$
$$= \sup\{\frac{1}{2} \lambda_{\max}(A_0(t) + A_0^T(t))\}$$
$$= \sup_{t \in \Re^+}(1 - 2e^t)$$
$$= 0.5.$$

Also, here $\|A\| = e^{-0.2}/40$.

In the present case, the Liapunov Equation is given by

$$\dot{P}(t)+[A^T_0(t)+A_0(t)]P(t)+2I = 0 \text{ (as } m=2).$$

$$\Rightarrow \dot{P}(t)+\begin{bmatrix} 1-2e^t & 0 \\ 0 & 1-2e^t \end{bmatrix}P(t)+\begin{bmatrix} 2 & 0 \\ 0 & 2 \end{bmatrix}=\begin{bmatrix} 0 & 0 \\ 0 & 0 \end{bmatrix}. \tag{6.11}$$

Here, $P(t)$ is a 2×2 matrix satisfying the matrix equation (6.11).

The matrix $P(t)$ can be obtained as $P(t) = \begin{bmatrix} e^{-t} & 0 \\ 0 & e^{-t} \end{bmatrix}$.

So, from definition, we have $P(t)+I = \begin{bmatrix} 1+e^{-t} & 0 \\ 0 & 1+e^{-t} \end{bmatrix}$ and so

$$\|P_I\| = \sup_{t \in \Re^+}(\|P(t)+I\|)$$
$$= \sqrt{\sup_{t \in \Re^+}\{(1+e^{-t})^2\}}$$
$$= 2.$$

Also $h=\max\{h_1, h_2\}=1$. Thus, the rate of convergence is given by the solution of the inequality (6.10), as $0.5 + 2\alpha + (e^{-0.2}/40)^2 e^{2\alpha} \leq 0$. The value of $\alpha$ is found to be 0.2.

Hence all the conditions of Theorem 6.3 are satisfied. Thus, the delay system in this case is 0.2-stable. This roughly means that under suitable conditions the delay system can be made to converge exponentially with a rate 0.2.

**Example 6.7:** Consider the linear autonomous delay system

$$\dot{x}(t) = A_0 x(t) + A_1 x(t-0.5) + A_2 x(t-1) \; , \; t \in R^+$$

with any given initial function $\phi(t) \in C\left([-1,0], R^2\right)$ and

$$A_0 = \begin{bmatrix} -2 & 0.5 \\ -1 & -4 \end{bmatrix}, \; A_1 = A_2 = e^{-0.4}\begin{bmatrix} 1/3 & 0 \\ 0 & 1/3 \end{bmatrix}.$$

Here, we have the values $m=2$, $h_1=0.5$, $h_2=1$.

Also, as usual $\eta(A_0) = \sup_{t \in \Re^+}\eta(A_0(t)) = -3 + 0.5\sqrt{4.25}$, $\|A\|^2 = \dfrac{e^{-0.8}}{9}$. The algebraic Liapunov Equation is given by: $A^T_0 P + PA_0 + 2I = 0$.

So, $\begin{bmatrix} -2 & -1 \\ 0.5 & -4 \end{bmatrix}P + P\begin{bmatrix} -2 & 0.5 \\ -1 & -4 \end{bmatrix}+\begin{bmatrix} 2 & 0 \\ 0 & 2 \end{bmatrix}=\begin{bmatrix} 0 & 0 \\ 0 & 0 \end{bmatrix}$. The solution of this equation is given by $P = \begin{bmatrix} 0.5 & 0 \\ 0 & 0.25 \end{bmatrix}$. So, we have here $P+I = \begin{bmatrix} 1.5 & 0 \\ 0 & 1.25 \end{bmatrix}$. Thus $\|P_I\| = \|P+I\| = 1.5$.

Now, as before we can find the rate of convergence as *α*=0.4. Hence the linear autonomous delay system is *0.4-stable* in this case.

### 6.4 An Important Observation

An important feature of usually viewing time-varying systems and understanding their behavior is to be precise about the functional dependencies. We have to be sure about the correct functional relationship among the functions involved in the system. For example, $f(x) = x^3$ is an increasing function of *x* for all $x \in R$. But, if we view the same function on the system trajectories given by $x(t) = 1/t$, $t > 0$, then as $f(x(t)) = 1/t^3$, $t > 0$, so *f*(*x*(*t*)) is decreasing function of *t* for all *t*>0. Similarly, if we consider *f*(*t*, *x*)=*t* + *x*, then it is increasing in both *t* and *x*, but along the trajectory *x*(*t*) =−2*t* it appears to be decreasing function of *t*.

Let us consider another example as

$$\left. \begin{array}{l} f_1(x_1) = x_1^2, x_1(t) = t^3 \\ f_2(x_2) = x_2^3, x_2(t) = t^2 \end{array} \right\}$$

Both leading to the same function of time $f_1(x_1(t)) = f_2(x_2(t)) = t^6$, but the underlying structures of *x* and *t* are completely of different nature. In control problems, it is very important to carefully distinguish between the functional dependencies of the variables involved. Thus, while dealing with the systems consisting of various components one should not be confused with the relationship among them or otherwise often misleading results may come into play.

### Concluding Remarks:

The examples and theorems we have seen in this chapter deal with the important features of the exponential and *α*-stability. These are very important in practical problems, where we are not satisfied only with the asymptotic stability of a system but also want to know the rate, i.e. how fast the system approaches the point of equilibrium. An exponentially stable system with a fast rate of convergence is of course more desirable in physical situations. Also, we have seen that the stability of non-autonomous systems largely depend on the functional relationships of the variables, such as time and state. Thus, we have to be clear about this before jumping into any conclusions regarding them.

*Chapter-7*

# LIAPUNOV'S STABILITY THEORY FOR NON-AUTONOMOUS SYSTEMS

## 7.1 Preliminary Introduction

In Chapter 4, we have studied Liapunov's concept and formulation of the stability theory for time-invariant systems. The main aim of that chapter was to introduce and implement the celebrated Liapunov's stability theorem which is the consequence of the direct Liapunov's method. In this regard, we have seen the crucial significance of a Liapunov's function in assessing stability or instability of the concerned autonomous system. As mentioned in that chapter, Liapunov's direct method is applicable to both autonomous as well as non-autonomous dynamical systems. The present chapter is destined to develop the Liapunov's stability theory for non-autonomous systems. At first, we recall and necessarily define some important notions which will be used throughout the chapter.

- Let $V(x)$ be an autonomous continuously differentiable function of the $n$ variables $x_1$, $x_2,...,x_n$ defined in $R^n$. Then $V$ is said to be *positive definite* in a region $U$ of $R^n$ containing the origin, if:
    a) $V(0) = 0$ and
    b) $V(x) \geq 0$ for all $x \in U$ and $x \neq 0$.
- Also $V(x)$ is said to be *positive semi-definite* if $V(x) \geq 0$ for $x \in U$ and $x \neq 0$.
- Just reversing the direction of the inequality signs in the above two definitions we can define a *negative definite* and n*egative semi-definite* function in some region $U$ of $R^n$.

Now, here we notice that if in the domain $U$ the function $V(x)$ keeps only one sign but can become zero at some point other than origin then it is correspondingly said to be positive or negative semi-definite. On the other hand if a semi-definite function vanishes only at the origin $x=0$, then it is called a definite function.

Let us now define definite and semi-definite time-varying functions.

Let $V(x,t): D \times R^+ \to R$ be a non-autonomous function of $x$ and $t$ such that $V(x,t)$ is continuous and has continuous partial derivatives with respect to all of its arguments and $D$ is an open set containing the origin.

- $V(x,t)$ is said to be *positive semi-definite* in $D$ if:
    a) $V(0,t)=0 \quad \forall t \in R^+$.
    b) $V(x,t) \geq 0$ for all $x \in D$ and $x \neq 0$.

Similarly, we can define negative definiteness also.

- $V(x,t)$ is said to be *decrescent* in $D$ if there exists a positive definite autonomous function $W_1(x)$ such that

$$|V(x,t)| \leq W_1(x) \, \forall x \in D.$$

This means that every time-invariant positive definite function is decrescent.

- $V(x,t)$ is said to be *radially* unbounded if $V(x,t) \to \infty$ as $\|x\| \to \infty$. This means that given $M, \exists N > 0$, such that $V(x,t) > M$ for all $t$ provided that $\|x\| > N$.

**NOTE:** A function $V(x,t)$ which explicitly depends on time $t$ is positive definite iff there exists a positive definite function $W(x)$, independent of time $t$, such that $V(x,t) \geq W(x) \, \forall x \in D$. In a similar manner, $V$ is negative definite iff $-V(x,t) \geq W(x) \, \forall x \in D$.

$V(x,t)$ is both positive definite and decrescent iff there exist positive definite time-invariant functions $W_1(x)$ and $W_2(x)$ such that

$$W_1(x) \leq V(x,t) \leq W_2(x) \, \forall x \in D.$$

For a positive definite function $V(x,t)$ that depends explicitly on time $t$, the existence of a lower bound function may be demonstrated geometrically. In the space $V$, $x_1, x_2,...,x_n$, we can construct the two surface $W=W(x)$ as well as the surface $V(x,t)$ at time $t$ in the phase space.

As $t$ changes, the surface $V(x,t)$ will also change but will never go below the lower bound surface $W=W(x)$, as shown in Fig. 7.1.

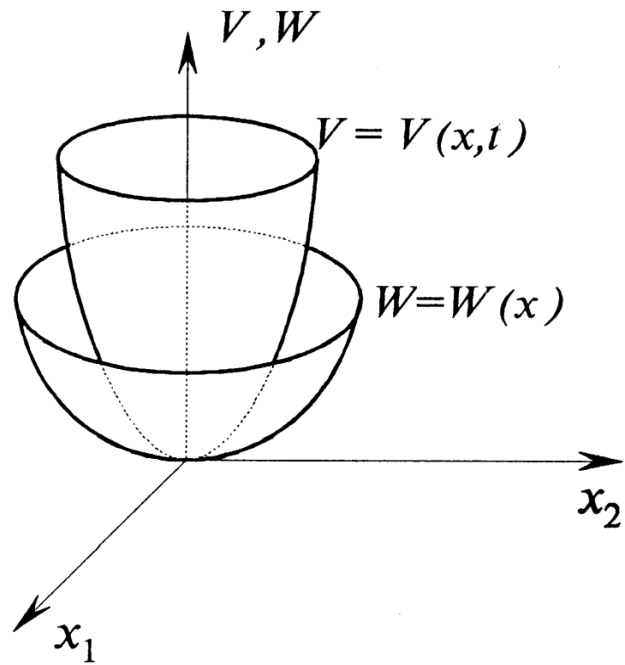

**Fig 7.1: The surfaces $W=W(x)$ and $V(x,t)$**

## 7.2 The Generalized Sylvester's Criterion

Let $V(x,t)$ be a quadratic function, such as $V(x,t) = \dfrac{1}{2}\sum_{k=1}^{n}\sum_{j=1}^{n}\alpha_{kj}x_k x_j$, where $\alpha_{kj}$ are various functions of time $t$ and the variables $x_j$. We assume that $V(x,t)$ is defined for all the real values of time $t$ and the variables $x_1, x_2,...,x_n$ which satisfy the relations

$$\sum x_j^2 < \mu,\ t \geq t_0 \tag{7.1}$$

If in the domain of the inequality (7.1), for a sufficiently large $t_0$ and a sufficiently small $\mu$, all principal diagonal minors of the matrix

$$\begin{bmatrix} \alpha_{11}(x,t) & \cdots & \cdots & \alpha_{1n}(x,t) \\ \cdots & \cdots & \cdots & \cdots \\ \cdots & \cdots & \cdots & \cdots \\ \alpha_{n1}(x,t) & \cdots & \cdots & \alpha_{nn}(x,t) \end{bmatrix}$$

satisfy the *generalized Sylvester criterion*, given by

$$\Delta_1 = \alpha_{11} \geq \delta_1 > 0, \ldots, \Delta_n = \begin{vmatrix} \alpha_{11} & \cdots & \cdots & \alpha_{1n} \\ \cdots & \cdots & \cdots & \cdots \\ \cdots & \cdots & \cdots & \cdots \\ \alpha_{n1} & \cdots & \cdots & \alpha_{nn} \end{vmatrix} \geq \delta_n > 0, \tag{7.2}$$

Where $\delta_1, \delta_2,\ldots,\delta_n$ are positive constants, then (according to Liapunov), the function $V(x,t)$, as defined above will be positive definite.

Now, we give some important mathematical examples in order to illustrate the concepts, described so far.

**Example 7.1:** Let us consider the following non-autonomous function

$$V(x,t) = t(x_1^2 + x_2^2) - 2\cos t.x_1 x_2.$$

Here the co-efficient matrix is given by

$$\begin{bmatrix} t & -\cos t \\ \cos t & t \end{bmatrix}.$$

The principal diagonal minors are as $\Delta_1 = t$, $\Delta_2 = t^2 - \cos^2 t$. Now if we assume that $t_0 = 1$, then for all $t \geq 1$, we get,

$$\Delta_1 \geq 1 > 0,\ \Delta_2 = t1 - \cos^2 1 \approx 0.71 > 0.$$

So, here the generalized Sylvester criterion is fulfilled and hence according to Liapunov, the given function is positive definite.

**Example 7.2:** Consider the time-varying function, given by

$$V(x,t) = [1 - a\cos\{(x_1^2 + x_2^2)t\}]x_1^2 + 2a\sin[(x_1^2 + x_2^2)t]x_1 x_2 + [1 + a\cos\{(x_1^2 + x_2^2)t\}]x_2^2.$$

Here the matrix of coefficients is given by

$$\begin{bmatrix} 1 - a\cos\{(x_1^2 + x_2^2)t\} & a\sin\{(x_1^2 + x_2^2)t\} \\ a\sin\{(x_1^2 + x_2^2)t\} & 1 + a\cos\{(x_1^2 + x_2^2)t\} \end{bmatrix}.$$

Now, the principal diagonal minors are found as $\Delta_1 = 1 - a\cos(x_1^2 + x_2^2)t\}$, $\Delta_2 = 1 - a^2$.

From above it is clear that for all $t$, $x_1$, $x_2$, we have $\Delta_1 \geq 1 - |a|$, $\Delta_2 = 1 - a^2$. Thus, we see that according to Liapunov, for all $|a| < 1$, the given function $V(x,t)$ is positive definite.

**Example 7.3:** Let $V(x,t) = (x_1^2 + x_2^2)e^{-\alpha t}$, $\alpha > 0$. Here, we have

$$V(0,t) = 0e^{-\alpha t} = 0$$
$$V(x,t) > 0 \ \forall x \neq 0, \ t \in R.$$

Thus $V(x,t)$ in this case is positive semi-definite but not positive definite.

**Example 7.4:** Let us now take the non-autonomous function

$$V(x,t) = \frac{(x_1^2 + x_2^2)(t^2 + 1)}{(x_1^2 + 2)}$$
$$= W(x)(t^2 + 1) \text{ (say)}$$

Thus here $V(x,t) \geq W(x) > 0 \forall x \in R^2$ and so $V(x,t)$ is positive definite from definition. Also we see that, $V(x,t) \geq W(x) > 0 \forall x \in R^2$. Thus, we cannot find a positive definite function $W_1$ such that $|V(x,t)| < W_1(x) \ \forall x \in R^2$ and so $V(x,t)$ is not decrescent. Also it is not radially unbounded as here $V(x,t)$ does not tends to $\infty$ as $|x| \to \infty$.

**Example 7.5:** As an opposition of the previous example, we can see that the function $V(x,t) = (x_1^2 + x_2^2)(t^2 + 1)$ is positive definite, radially unbounded and decrescent.

**Example 7.6:** The function $V(x,t) = \dfrac{(x_1^2 + x_2^2)}{(x_1^2 + 1)}$ is positive definite. It is independent of time and so is clearly decrescent. Also as we can easily see that this function is not radially unbounded.

## 7.3 Limited Systems

We define here another important category of functions. If, under the conditions of Sect. 7.2, all the magnitudes of $|V|$ are less than some positive bound then the function $V$ is called *limited*. For the sake of continuity, for a sufficiently small $\mu$, any function $V$ that does not depend on the time $t$ is such a limited function. Also by virtue of continuity, every function $V$ that does not depend on time has an infinitely small upper limit. On the other hand, the functions that depend on time may not have such an infinitely small upper limit, even when they are limited.

Roughly speaking, by an infinitely small upper limit, we mean that for any $t \geq t_0$, the absolute value of $V(x,t)$ can be made arbitrarily small only by decreasing the absolute values of all $x_j$. As another illustrative example, we can observe that the following two time-varying functions $\sin^2[(x_1^2 + x_2^2 + \ldots + x_n^2)t]$ and $(x_1^2 + x_2^2 + \ldots + x_n^2)\sin^2 t$ are limited as well as positive definite but neither of these two functions is semi-definite and only the second function has an infinitely small upper limit. On the other hand, the function $t(x_1^2 + x_2^2) - 2\cos t . x_1 x_2$ is positive definite but is not limited. Hence it has no infinitely small upper limit.

With the above definitions and concepts, we are now going to discuss the main theorems and consequences of the Liapunov's direct method for stability of non-autonomous systems. Some of these upcoming theorems can be stated and proved in a manner very similar to the corresponding results for autonomous systems.

## 7.4 Liapunov's Direct Method for Non-autonomous Systems

Let us consider the non-autonomous system, given by

$$\dot{x}(t) = f(x(t),t), \; x(t_0) = x_0, \; x \in R^n \tag{7.3}$$

Here $f$ is piecewise continuous in $t$ and locally Lipschitz in $x$ on $D \times [t_0, \alpha)$. $D$ is an open set containing the equilibrium point $x=0$ of the system. Also we assume that the initial time $t_0 \geq 0$. Now, we state the following theorems regarding the stability of this system:

**Theorem 7.1: (Liapunov's Stability theorem)**

If in a neighborhood $D$ of the equilibrium point $x=0$ of the system (7.3) there exists a function $V(x,t): D \times [0,\infty) \to R$, such that

a) $V(x, t)$ is positive definite in $D$.
b) The total differential of $V(x,t)$ with respect to time $t$ is negative semi definite in $D$ along any solution of the system (5.3), where

$$\dot{V} = \frac{\partial V}{\partial x_1} f_1 + \frac{\partial V}{\partial x_2} f_2 + \ldots\ldots\ldots + \frac{\partial V}{\partial x_n} f_n + \frac{\partial V}{\partial t}$$

$$= \frac{\partial V}{\partial x} f(x,t) + \frac{\partial V}{\partial t}.$$

Then the equilibrium point *x*=0 is stable. If *V*(*x*, *t*) is also decrescent then the origin is uniformly stable for the system.

**Theorem 7.2: (Liapunov's theorem for asymptotic stability)**
If in a neighborhood *D* of the equilibrium state *x*=0 of the system (7.3) there exists a function $V(x,t): D \times [0,\infty) \to R$, such that
  a) *V*(*x*,*t*) is positive definite and decrescent, i.e. there exist two positive definite functions $W_1(x)$ and $W_2(x)$ in *D* such that

  $$W_1(x) \leq V(x,t) \leq W_2(x) \quad \forall x \in D.$$

  b) The derivative $\dot{V}(x,t)$ is negative definite in *D*, i.e. there exist a function $W_3(x)$ in *D* such that

  $$\dot{V}(x,t) = \frac{\partial V}{\partial x} f(x,t) + \frac{\partial V}{\partial t} \leq -W_3(x) \quad \forall t \geq t_0, \; x \in D.$$

Then, *x*=0 is uniformly asymptotically stable. Moreover if the function *V*(*x*,*t*) is also radially unbounded then the origin *x*=0 is a globally uniformly asymptotically stable equilibrium point.

**Theorem 7.3: (Liapunov's theorem for exponential stability)**
Suppose that all the conditions of the Theorem 7.2 are satisfied and also in addition assume that there exist positive constants $K_1$, $K_2$ and $K_3$ such that

$$\left. \begin{array}{l} K_1 \|x\|^p \leq V(x,t) \leq K_2 \|x\|^p \\ \dot{V}(x,t) \leq -K_3 \|x\|^p \end{array} \right\}$$

Then the origin is exponentially stable. Further, if the above conditions hold globally then origin is globally exponentially stable.

**NOTE: Some important remarks**
The foregoing theorems give the sufficient criteria for asymptotic, uniform and exponential stability. However, these are by no means the necessary conditions. The function *V*(*x*,*t*) in all the above cases is called the *Liapunov Function*. If for a system, we can find one such function then we can immediately guarantee its stability. However if we cannot find any such function it does not mean that the system is unstable. Thus, failure of finding a Liapunov function to satisfy the conditions for stability or asymptotic stability does not mean that the equilibrium is not stable or not asymptotically stable; rather, it only means that such properties cannot be established using this particular Liapunov function candidate. There are no generally applicable methods for finding Liapunov functions. Different trial and error or some mathematical or physical insights according to the problem are often used. A useful procedure for this purpose is known as the *variable gradient method*.

Let us now see some of the applications of the direct method:

**Example 7.7:** Consider the scalar time-varying system, described by

$$\dot{x}(t) = -x^3 + \frac{x^3}{2}\sin t, \quad x(t_0) = x_0 \tag{7.4}$$

Here, $x=0$ is the only point of equilibrium. To study the stability of the above time-varying system at this point.

Let us use the following as a Liapunov function candidate: $V(x,t) = \frac{x^2}{2}$. Now, we obtain the time derivative of $V(x,t)$ along the trajectories of the system:

$$\begin{aligned}\dot{V}(x,t) &= \frac{\partial V}{\partial x}f(x,t) + \frac{\partial V}{\partial t} \\ &= \frac{\partial V}{\partial x}(-x^3 + \frac{x^3}{2}\sin t) \\ &= x(-x^3 + \frac{x^3}{2}\sin t) \\ &= -x^4(1 - \frac{1}{2}\sin t)\end{aligned}$$

Now, we choose $W_1(x)=W_2(x)=V(x)$ and $W_3(x)=ax^4$, where $a<0{:}5$. Then, all the conditions of the Theorem 7.2 are satisfied, globally. Hence, in this case, the origin $x=0$ is globally uniformly asymptotically stable.

**Example 7.8:** Let us consider the following system

$$\left.\begin{aligned}\dot{x}_1(t) &= -x_1 - e^{-2t}x_2 = f_1 \\ \dot{x}_2(t) &= x_1 - x_2 \quad\quad = f_2\end{aligned}\right\} \tag{7.5}$$

Here, the origin (0,0) is the only equilibrium point. To study the stability at the origin, let us choose as a Liapunov function candidate

$$V(x,t) = x_1^2 + (1 + e^{-2t})x_2^2.$$

Also let $W_1(x) = x_1^2 + x_2^2$, $W_2(x) = x_1^2 + 2x_2^2$. Then, we can see that

$$W_1(x) \leq V(x,t) \leq W_2(x) \quad \forall x_1, x_2 \tag{7.6}$$

Thus, from the above inequality we conclude the followings:

  a) $V(x,t)$ is positive definite, since $V(x,t) \geq W_1(x)$, with $W_1$ positive definite in $R^2$.
  b) $V(x,t)$ is decrescent, since $V(x,t) \leq W_2(x)$, with $W_2$ also positive definite in $R^2$.

Now we have to find the time derivative of $V(x,t)$. We see that

$$\dot{V}(x,t) = \frac{\partial V}{\partial x} f(x,t) + \frac{\partial V}{\partial t}$$
$$= 2x_1(-x_1 - e^{-2t}x_2) + 2x_2(1 + e^{-2t})(x_1 - x_2)$$
$$= -2[x_1^2 - x_1 x_2 + x_2^2(1 + 2e^{-2t})]$$
$$\leq -2[x_1^2 - x_1 x_2 + 3x_2^2]$$

Let $W_3(x) = 2[x_1^2 - x_1 x_2 + 3x_2^2]$. Then, we see that the coefficient matrix for $W_3$ is $\begin{bmatrix} 2 & -1 \\ -1 & 6 \end{bmatrix}$. It has the principal diagonal minors as $\Delta_1 = 2, \Delta_2 = 11$ which are both positive. Thus, the function $W_3(x)$ is positive definite and so, we have the time derivative of $V(x,t)$ is negative definite.

Hence all the conditions of the Theorem 7.2 are satisfied and so, in this case the origin is a globally asymptotically stable equilibrium point.

### NOTE: Concluding remarks

In this chapter, we have seen the technique of Liapunov's direct method to analyze stability properties of time-varying systems. This method is widely used in practical problems due to its amazing ability. The two examples, we have taken show that the construction of Liapunov function depends on the nature of the particular problem and other suitable conditions.

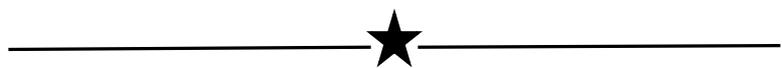

*Chapter-8*

# PERIODIC AND DISCRETE TIME STABILITY

### 8.1 Linear Systems with Periodic Coefficients

Stability investigation of various practical systems often reduces to the analysis of linear differential equations with periodic coefficients. So, it is important to discuss about the stability criteria of time-varying systems with periodic components. In matrix form such a system can be expressed as follows

$$\dot{x} = P(t)x \tag{8.1}$$

where, $x = \begin{bmatrix} x_1(t) \\ x_2(t) \\ \cdots \\ \cdots \\ \cdots \\ x_n(t) \end{bmatrix}$ is a column vector and $P(t) = \begin{bmatrix} p_{11}(t) & \cdots & \cdots & \cdots & p_{1n}(t) \\ p_{21}(t) & \cdots & \cdots & \cdots & p_{2n}(t) \\ \cdots & \cdots & \cdots & \cdots & \cdots \\ \cdots & \cdots & \cdots & \cdots & \cdots \\ p_{2n}(t) & \cdots & \cdots & \cdots & p_{2n}(t) \end{bmatrix}$ is a square matrix. We assume that all the elements of the matrix $P(t)$ and thus the matrix itself is a periodic function of time $t$ with a fixed period $T$. We express this as $P(t+T)=P(t)$ for all time $t$. Now we introduce some important terms which will be used here.

The set of $n$ linearly independent solutions of the system (8.1), given by

$x_1 = \begin{bmatrix} x_{11} \\ x_{21} \\ \cdots \\ \cdots \\ \cdots \\ x_{n1} \end{bmatrix}$, $x_2 = \begin{bmatrix} x_{12} \\ x_{22} \\ \cdots \\ \cdots \\ \cdots \\ x_{n2} \end{bmatrix}$, ..., $x_n = \begin{bmatrix} x_{1n} \\ x_{2n} \\ \cdots \\ \cdots \\ \cdots \\ x_{nn} \end{bmatrix}$ is called the *fundamental system of solutions* of the system and the matrix $X(t)=(x_1, x_2,..., x_n)$ is called the *fundamental matrix*. Here, in $x_{kj}$ the first index denotes the index of the function and the second denotes the index of solution. The general solution of the system (8.1) is will then be

$$x(t) = c_1 x_1 + c_2 x_2 + \cdots + c_n x_n$$

where, $c_i$'s are constants.

In matrix form, the solution can be written as, $x(t)=X(t)C$ where C is the column vector consisting of the elements $c_1, c_2,...,c_n$.

Now without loss of generality we may assume that the fundamental system of solutions satisfies the following conditions:

$$x_{kj} = \begin{cases} 1, k = j \\ 0, k \neq j \end{cases}$$

or in matrix form we can write this as, X(0)=I where I is the $n \times n$ identity matrix. We denote the determinant of the fundamental matrix by $\Delta(t) = \det X(t)$. Thus, using the conditions satisfied by the fundamental systems of solutions, we have

$$\Delta(0) = \det X(0) = \det I = 1 \tag{8.2}$$

Now the *Liouville formula* for differential equations is given by

$$\Delta(t) = \Delta(0) e^{\int_0^t (p_{11} + p_{22} + \ldots + p_{nn}) dt}.$$

Using this formula in this case, for $t=T$, we have,

$$\Delta(T) = \Delta(0) e^{\int_0^T (p_{11} + p_{22} + \ldots + p_{nn}) dt} \tag{8.3}$$

Any solution of the system (8.1) can be obtained from the general solution by a proper choice of the column matrix C, as

$$X_k(t+k) = X(t) A_k \ \forall k = 1, 2, \cdots, n.$$

So, the fundamental matrix which corresponds to the solutions $x_1(t+T)$, $x_2(t+T)$, ... ,$x_n(t+T)$ is given as

$$X(t+T) = X(t) A \tag{8.4}$$

where, A is the constant matrix, given by:

$$A = (A_1, A_2, \ldots, A_n)$$
$$= \begin{bmatrix} a_{11} & a_{12} & \ldots & \ldots & a_{1n} \\ a_{21} & a_{22} & \ldots & \ldots & a_{2n} \\ \ldots & \ldots & \ldots & \ldots & \ldots \\ \ldots & \ldots & \ldots & \ldots & \ldots \\ a_{n1} & a_{n2} & \ldots & \ldots & a_{nn} \end{bmatrix}.$$

As the fundamental system of solutions satisfy the initial conditions given before, so assuming $t=0$, we get from (8.4)

$$X(t) = X(0) A = IA = A \tag{8.5}$$

The above equation is very useful to determine the matrix A when the matrix X($t$) is known. Now, we state the following crucial result:

It is proved that for the periodic system (8.1), there always exist a solution $x(t)$ such that $x(t+T)=\rho x(t)$, where $\rho$ is some constant. Such a solution is called an **Orthogonal** or a **Normal** solution.

From the above result, it is clear that if a normal solution exists then there should exist a constant column matrix β where for which the equation $x(t)=X(t)\beta$ is valid. The constant matrix $\beta$ is given by

$$\beta = \begin{bmatrix} \beta_1 \\ \beta_1 \\ \ldots \\ \ldots \\ \ldots \\ \beta_n \end{bmatrix}.$$

Now by our assumption, $x(t)$ satisfies the equation $x(t+T)=\rho x(t)$ and so we consider the equation $x(t+T)=X(t+T)\beta$ from which we have $X(t+T)=X(t+T)\beta$. Using (8.4), we have $X(t)A\beta= \rho X(t)\beta$ or $X(t)(A-\rho I)\beta=0$. As this equation is satisfied for all $t$, so we have $(A-\rho I)\beta=0$. Now, this above matrix equation has a nontrivial solution in β, if and only if we have

$$|A-\rho I| = \begin{vmatrix} a_{11}-\rho & a_{12} & \ldots & \ldots & a_{1n} \\ a_{21} & a_{22}-\rho & \ldots & \ldots & a_{2n} \\ \ldots & \ldots & \ldots & \ldots & \ldots \\ \ldots & \ldots & \ldots & \ldots & \ldots \\ a_{n1} & a_{n2} & \ldots & \ldots & a_{nn}-\rho \end{vmatrix} = 0 \tag{8.6}$$

The matrix equation (8.6) is known as the *characteristic equation*. Each root $\rho_k$ of this characteristic equation corresponds to a specific solution $x_k(t)$. As a result, we get a set of $n$ normal solutions $x_1(t), x_2(t), ..., x_n(t)$ satisfying the normal equation.

## 8.2 Stability Analysis

After obtaining the characteristic equation (8.6), we now proceed to investigate the stability of the periodic system (8.1) at the origin. We state below the following conditions of stability:

➢ If the moduli of all the roots of the characteristic equation (8.6) are less than unity then the periodic system (8.1) is asymptotically stable at the origin.

➢ If even one of the roots of the characteristic equation has a modulus larger than unity then the periodic system (8.1) is asymptotically unstable at the origin.

> If the moduli of some of the roots of the characteristic equation are equal to unity while the remaining ones are less than unity, then the motion is stable but not asymptotically at the origin.

Thus, to investigate the stability of periodic systems, the most important thing is to construct the characteristic equation and find its roots.

**NOTE: An important observation**

Often it is quite difficult to find even one fundamental matrix X(*t*) which is very much essential to apply the criteria for stability because we have X(*T*)=A. In such cases, we use the numerical analysis or other suitable method to estimate at least one fundamental matrix. After that the characteristic roots are approximated.

One check for the accuracy of the characteristic roots is given by:

$$\rho_1 \rho_2 \ldots \rho_n = e^{\int_0^T (p_{11}+p_{22}+\ldots+p_{nn})dt}.$$

This is one major drawback with stability of periodic motions in practical problems.

**Example 8.1:** Let us consider the following non-autonomous system

$$\begin{aligned}\dot{x}_1 &= -x_1 + \sin t.x_2 \\ \dot{x}_2 &= \cos t.x_1 - x_2 - \sin t.x_3 \\ \dot{x} &= \cos t.x_2 - x_3\end{aligned}\quad (8.7)$$

In matrix form, the system can be written as $\dot{x} = P(t)x$ where the coefficient matrix $P(t)$ is given by

$$P(t) = \begin{bmatrix} -1 & \sin t & 0 \\ \cos t & -1 & -\sin t \\ 0 & \cos t & -1 \end{bmatrix}.$$

This is a periodic system with period $2n\pi$. We want to investigate the stability of the system at origin in the interval [0,2π], where *t*=0 is the initial time. Now, here we use the method of numerical integration to find three linearly independent solutions of the system in the interval [0,2π]. Then, we can determine X(*t*), where X(0)=I. So we have X(2π)=A. Then, as usual we can find the characteristic roots as $\rho_1 = 2.566519.10^{-5}, \rho_{2,3} = 0.008405 \pm 0.013532i$. So, we have

$$e^{\int_0^T (p_{11}+p_{22}+\ldots+p_{nn})dt}$$
$$= e^{\int_0^{2\pi} (-3)dt}$$
$$= e^{-6\pi} = 6.512412 \times 10^{-9}$$

Also, $\rho_1\rho_2\rho_3 = 6.512428\times10^{-9}$. Thus, we have done a good approximation of the three characteristic roots. Since, all moduli of all the characteristic roots are less than unity, so the system is asymptotically stable at the origin for all $t>0$.

### 8.3 Introduction to Discrete Time Stability

As a contrary to continuous systems we can also have the discrete time systems, which are frequently encountered in practical problems. In discrete time systems, the state variables undergo discrete or finite jumps with the change of time $t$. We mathematically define a discrete time system as

$$x(k+1) = f(x(k), k) \tag{8.8}$$

where, $k \in Z^+$, $x(k) \in R^n$ and $f: R^n \times Z \to R^n$.

Discrete-time systems $\Sigma_d$: $u(k) \to x(k)$ may originate by *sampling* some continuous time system $\Sigma$: $u(t) \to x(t)$. We use the scheme, shown in Fig. 8.1.

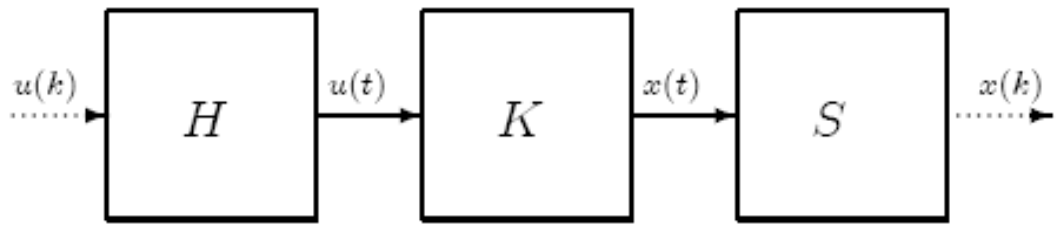

**Fig 8.1: Discrete-time system $\Sigma_d$**

In Fig. 8.1, we have
- ➤ *S: Sampler.* Reads $x(t)$ every $T$ seconds; $x(k)=x(kT)$.
- ➤ $\Sigma$: *The appropriate plant.* For a given $u$, $\Sigma$: $u \to x$ determines $x(t)$ by solving the equation $\dot{x} = f(x, u)$.
- ➤ *H: Hold device.* It converts $u(k)$ into $u(t)$ (continuous-time) as
  $u(t) = u(k)$ for all $kT \leq t < (k+1)T$.

Exact conversion of a continuous system to discrete time system is often impossible. There are several methods to construct approximate models. The simplest is the so-called *Euler approximation*. If $T$ is small, then

$$\dot{x} = \frac{dx}{dt} = \lim_{\Delta T \to 0} \frac{x(t+\Delta T) - x(t)}{\Delta T}$$
$$\approx \frac{x(t+T) - x(t)}{T}$$

Thus $\dot{x} = f(x, u)$ can be approximated by $x(k+1) \approx x(k) + T f[x(k), u(k)]$.

Let us see the corresponding differences in the diagram below:

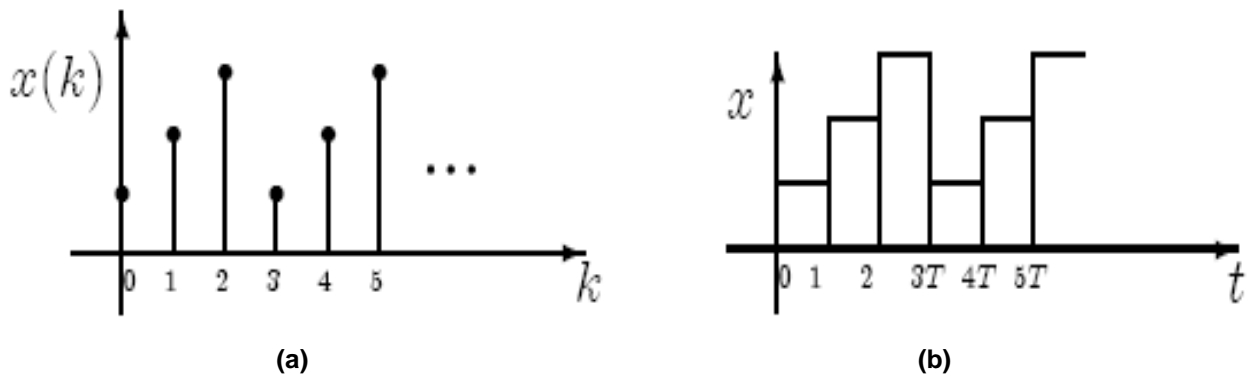

(a)  (b)

Fig 8.2: Difference between continuous and discrete systems: (a) continuous, (b) discrete

## 8.4 Stability of Discrete Time Systems

The stability analysis of discrete systems is similar to that of a continuous system in the sense that here derivative of $x$, i.e. $\dot{x}$ is replaced or approximated by the discrete difference term $\Delta x = x(k+1) - x(k)$. Thus, for the system (8.8), the point $x=x^e$ is an equilibrium point if $\Delta x^e = 0$, $k \geq k_0 > 0$, $k_0$ being the initial state. Similarly, we can define the stability at an equilibrium point.

As example, the equilibrium point $x=0$ of the system (8.8) is said to be *stable* at the state $k_0$ if for any given $\varepsilon>0$, there exists a real number $\delta=\delta(\varepsilon,k_0)>0$, such that $\|x(0)\| \leq \delta \Rightarrow \|x(k)\| \leq \varepsilon$ $\forall$ $k \geq k_0 > 0$. Also, exactly in a similar manner we can restate the stability theorems of direct method used in case of continuous systems. Let us now see one of the stability theorems here.

**Theorem 8.1:** If in a neighborhood $D$ of the equilibrium state $x=0$ of the system (8.8) there exists a function $V(x,k)$: $D \times Z^+ \to R$, such that
  a) $V(x, k)$ is positive definite in $D$.
  b) The rate of change $\Delta V(x,k) = V(x(k+1), k+1) - V(x,k)$ is negative semi definite in $D$ along any solution of the system (8.8),

Then the equilibrium state $x=0$ is stable. Moreover if $V(x, k)$ is also decrescent then the origin is uniformly stable for the system.

Other stability theorems follow for discrete time systems in a similar manner.

**Example 8.2:** Consider the following discrete time system:

$$\left. \begin{array}{l} x_1(k+1) = x_1(k) + x_2(k) \\ x_2(k+1) = ax_1^3(k) + \dfrac{1}{2}x_2(k) \end{array} \right\}$$

To study the stability at the equilibrium point origin, we consider the following (time independent) Liapunov function candidate

$$V(x(k)) = \frac{1}{2} x_1^2(k) + 2x_1(k)x_2(k) + 4x_2^2(k).$$

As in continuous case, this function *V(x)* can easily seen to be positive definite. Now, we need to find $\Delta V(x(k)) = V(x(k+1)) - V(x(k))$. We have

$$V(x(k+1)) = \frac{1}{2} x_1^2(k+1) + 2x_1(k+1)x_2(k+1) + 4x_2^2(k+1).$$

$$= \frac{1}{2}[x_1(k) + x_2(k)]^2 + 2[x_1(k) + x_2(k)][ax_1^3(k) + \frac{1}{2}x_2(k)] +$$

$$4[ax_1^3(k)\frac{1}{2}x_2(k)]^2.$$

$$V(x(k)) = \frac{1}{2} x_1^2(k) + 2x_1(k)x_2(k) + 4x_2^2(k).$$

From here after some trivial manipulations we conclude that

$$\Delta V(x(k)) = V(x(k+1)) - V(x(k))$$

$$= -\frac{3}{2} x_2^2(k) + 2ax_1^4(k) + 6ax_1^3(k)x_2(k) + 4a^2 x_1^6$$

Therefore we have the following cases of interest:
- *a*<0: in this case, ΔV (x) is negative definite in a neighborhood of the origin, and the origin is locally asymptotically stable (uniformly, since the system is autonomous).
- *a*=0: in this case $V(x(k)) = -\frac{3}{2} x_2^2(k) \leq 0$ and thus the origin is stable.

### NOTE: Concluding remarks

In this chapter we have studied the behaviors of periodic systems which very often occur in practical problems. Also, we have provided a brief overview of discrete time systems and we have seen that these can be handled in a similar manner as continuous cases with little bit changes in the underlying definitions and notations.

# SUMMARIZATION

Stability is a crucial requirement for a dynamical system which in a broad sense means that small disturbances either in system inputs or initial conditions do not lead to substantial changes in the overall behavior of the system. Of course, stability is the basic condition for a practically useful system; unstable systems just can be thought as useless. This book has meticulously dealt with the conceptualization and formulation of the stability theory for both autonomous as well as non-autonomous systems. It has been observed that stability of a linear autonomous system is very simple to determine. In fact, for a linear autonomous system, one only need to find the eigen values of the associated coefficient matrix and the signs of these eigen values are enough to indicate the nature of stability of the system. However, the situation becomes considerably difficult for nonlinear systems, as there is no general rule for assessing stability nature of these systems, unlike their linear counterparts. A remedial measure is the concept of linearization, discussed in Chapter 3. But this approach is limited to only a handful of cases and often it fails to give adequate information about the nature of stability.

Undoubtedly, the primary attraction of this book is the Liapunov's direct method which is so far the most suitable and perhaps the best method for determining nature of stability of a dynamical system. This method can be applied to both linear and nonlinear autonomous systems and even to non-autonomous systems as well. The only problem with the application of Liapunov's direct method is the selection of a suitable V-function whose derivative is either positive or negative definite. Unfortunately, till now, we have no general method for finding such a V-function for a given non-linear system. Formulation and implementation of Liapunov's direct method for stability analysis of time-independent systems has been studied with considerable details, covering various pros and cons in Chapter 4.

Now, when we turn to non-autonomous systems, the situation becomes entirely different. Often these systems behave in totally unexpected manners, as has been observed in Chapter 5 and as such one should be very cautious while dealing with them. Through suitable mathematical examples, we have seen that in case of a non-autonomous system, one should not use the usual results of autonomous systems just by finding some structural similarities because this often leads to completely wrong conclusion. However, we have also found that certain typical non-autonomous systems are in a sense equivalent to autonomous systems with their phase space extended to include the time component.

In Chapter 5, we have defined the various terms associated with the stability of time-varying systems. Although, some of the notations and terminologies were extended from the autonomous systems but most of them required fresh definitions. In spite of different approaches to deal with the stability of time-varying systems, here we have considered the Liapunov notion of stability which is based on the Euclidean norm. Then, we have defined some additional concepts such as convergence, global convergence, uniform stability etc. which are useful in the study of non-autonomous systems.

In the sixth chapter we have dealt with the exponential and $α$-stability which have significant practical applications. In many physical systems, we often want to know the rate at which a stable system approaches to its point of equilibrium. We have seen that there are some important results which give sufficient criteria for $α$-stability of a system with delay times. We have also noticed that an exponentially stable system is always uniformly stable but the converse is not always true.

The Liapunov's direct method which gives the sufficient conditions of stability of time-varying systems was studied in Chapter 7. This is so far the best method of stability analysis and widely applied to various practical systems. However, as stated earlier, the main and perhaps the most serious drawback of this method is that there is no well-defined rule to construct a suitable Liapunov function.

At last in Chapter 8, we have considered the analysis of discrete and periodic time stability. It has been noticed that the construction of fundamental matrix and finding its characteristic roots are necessary to deal with the periodic systems. Also, we have seen that the discrete time systems can be studied exactly in the same manner as continuous systems only by changing some definitions and introducing some more notations. Moreover, a continuous system can be approximated to a discrete system through a suitable discretization.

From our overall study, we ultimately conclude that it is relatively very much difficult to deal with the time-varying systems. The inclusion of the time component creates considerable complexities to a system. Still in spite of these difficulties we must have to consider them as they reflect the actual view of practical situations.

Stability theory is a vast subject area which is continuously growing with new innovations and concepts. As such, a comprehensive treatment of this subject cannot be provided in an introductory book, like this one. The primary aim of this book was to make the readers meticulously familiar with the notion of stability and its practical significance. Of course, this subject has lot much to explore and hence provides an excellent scope for future studies by the interested readers.